\pdfoutput=1

\documentclass[modern]{aastex62arxiv}
\usepackage{amsmath}
\usepackage{booktabs}
\usepackage{graphicx}

\newcommand{\asnow}{a_\mathrm{snow}}
\newcommand{\au}{\mathrm{au}}
\newcommand{\chisqred}{\chi^2_\mathrm{red}}

\newcommand{\diff}[2]{\mathrm{d}^{#1}#2}

\renewcommand{\deg}{\mathrm{deg}}
\newcommand{\DL}{D_\mathrm{L}}
\newcommand{\DS}{D_\mathrm{S}}
\newcommand{\eg}{e.g.}
\newcommand{\eq}[1]{equation~(\ref{#1})}
\newcommand{\Eq}[1]{Equation~(\ref{#1})}
\newcommand{\Fig}[1]{Figure~\ref{#1}}
\newcommand{\fbl}[1]{f_{\mathrm{b},#1}} 
\newcommand{\fsl}[1]{f_{\mathrm{s},#1}} 
\newcommand{\G}{G}

\newcommand{\ie}{i.e.}
\newcommand{\Is}{I_\mathrm{S}}
\newcommand{\Iso}{I_\mathrm{S,0}}

\newcommand{\kpc}{\mathrm{kpc}}
\newcommand{\mrm}[1]{\mathrm{#1}}

\newcommand{\mas}{\mathrm{mas}}
\newcommand{\meter}{\mathrm{m}}
\newcommand{\minu}{\mathrm{min}}
\newcommand{\Mjup}{M_\mathrm{J}}
\newcommand{\Msun}{M_\sun}
\newcommand{\Mearth}{M_\oplus}
\newcommand{\ML}{M}
\newcommand{\MLa}{M_1}
\newcommand{\MLb}{M_2}
\newcommand{\murelg}{\mu_\textrm{rel,G}}
\newcommand{\p}{\,.}

\newcommand{\Sec}[1]{Section~\ref{#1}}
\newcommand{\Tab}[1]{Table~\ref{#1}}
\newcommand{\thE}{\theta_\mathrm{E}}
\newcommand{\thS}{\theta_\star}
\newcommand{\tE}{t_\mathrm{E}}

\newcommand{\tS}{t_\star}

\newcommand{\thjd}{\mathrm{HJD^\prime}}
\newcommand{\uas}{\mu\mathrm{as}}

\renewcommand{\v}{\,,}

\newcommand{\yr}{\mathrm{yr}}

\newcommand{\murel}{\mu_\mathrm{rel}}

\newcommand\ltsima{$\; \buildrel <\over\sim \;$}
\newcommand\simlt{\lower.5ex\hbox{\ltsima}}
\newcommand\gtsima{$\; \buildrel >\over\sim \;$}
\newcommand\simgt{\lower.5ex\hbox{\gtsima}}

\newcommand{\Rmoa}{R_\mathrm{MOA}}

\newcommand{\Iogle}{I_\mathrm{OGLE}}

\newcommand{\RK}{R_\mathrm{MOA}}
\newcommand{\VK}{V_\mathrm{MOA}}

\newcommand{\ob}[2]{OGLE-20#1-BLG-#2}
\newcommand{\mb}[2]{MOA-20#1-BLG-#2}

\newcommand{\phntw}{\phn\phn}
\newcommand{\phnth}{\phn\phn\phn}
\newcommand{\phnfo}{\phn\phn\phn\phn}

\newcommand{\MLaval}{0.55\pm0.28}
\newcommand{\MLbval}{17.9^{+9.6}_{-8.8}}
\newcommand{\aperpval}{2.62^{+0.58}_{-0.60}}
\newcommand{\aval}{3.2^{+1.8}_{-0.8}}
\newcommand{\DLval}{6.7^{+1.0}_{-1.3}}
\newcommand{\thEval}{0.382^{+0.087}_{-0.076}}
\newcommand{\murelgval}{6.0\pm1.2}
\newcommand{\thSval}{0.784^{+0.093}_{-0.13}}
\newcommand{\JLvals}{22.0^{+1.3}_{-1.7}}
\newcommand{\JLvalss}{22.0^{+2.7}_{-2.6}}
\newcommand{\HLvals}{20.8^{+1.3}_{-1.6}}
\newcommand{\HLvalss}{20.8^{+2.7}_{-2.3}}
\newcommand{\KLvals}{20.3^{+1.3}_{-1.5}}
\newcommand{\KLvalss}{20.3^{+2.6}_{-2.2}}

\newcommand{\affmoaa}{Department of Earth and Space Science, Graduate School of Science, Osaka University, 1-1 Machikaneyama, Toyonaka, Osaka 560-0043, Japan}
\newcommand{\affmoab}{Institute for Space-Earth Environmental Research, Nagoya University, Furo-cho, Chikusa, Nagoya, Aichi 464-8601, Japan}

\newcommand{\affmoad}{Department of Physics, University of Auckland, Private Bag 92019, Auckland, New Zealand}
\newcommand{\affmoae}{Subaru Telescope Okayama Branch Office, National Astronomical Observatory of Japan, NINS, 3037-5 Honjo, Kamogata, Asakuchi, Okayama 719-0232, Japan}
\newcommand{\affmoaf}{Institute of Natural and Mathematical Sciences, Massey University, Private Bag 102904 North Shore Mail Centre, Auckland 0745, New Zealand}
\newcommand{\affmoag}{School of Chemical and Physical Sciences, Victoria University, Wellington, New Zealand}
\newcommand{\affmoah}{University of Canterbury Mt. John Observatory, P.O. Box 56, Lake Tekapo 8770, New Zealand}
\newcommand{\affmoai}{Department of Physics, Faculty of Science, Kyoto Sangyo University, 603-8555 Kyoto, Japan}
\newcommand{\affmoaj}{Institute of Space and Astronautical Science, Japan Aerospace Exploration Agency, 3-1-1 Yoshinodai, Chuo, Sagamihara, Kanagawa, 252-5210, Japan}

\newcommand{\affmoal}{Department of Astronomy, Graduate School of Science, The University of Tokyo, 7-3-1 Hongo, Bunkyo-ku, Tokyo 113-0033, Japan}
\newcommand{\affmoam}{National Astronomical Observatory of Japan, 2-21-1 Osawa, Mitaka, Tokyo 181-8588, Japan}
\newcommand{\affmoan}{Instituto de Astrof\'isica de Canarias, V\'ia L\'actea s/n, E-38205 La Laguna, Tenerife, Spain}

\received{2018 September 26}
\revised{2019 March 4}
\accepted{2019 March 26}
\published{2019 May 20}

\shorttitle{\footnotesize A Cold Neptune beyond the Snow Line in the Provisional \textit{WFIRST} Field}
\shortauthors{\footnotesize C. Ranc et al.}

\begin{document}

\title{\bfseries OGLE-2015-BLG-1670Lb: A Cold Neptune beyond the Snow Line in the Provisional \textit{WFIRST} Microlensing Survey Field}

\correspondingauthor{Cl{\'e}ment Ranc}
\email{clement.ranc@nasa.gov}

\author[0000-0003-2388-4534]{Cl{\'e}ment Ranc}%
\altaffiliation{The MOA Collaboration}
\affiliation{Astrophysics Science Division, NASA/Goddard Space Flight Center, Greenbelt, MD 20771, USA}

\author[0000-0001-8043-8413]{David P. Bennett}%
\altaffiliation{The MOA Collaboration}
\affiliation{Astrophysics Science Division, NASA/Goddard Space Flight Center, Greenbelt, MD 20771, USA}
\affiliation{Department of Astronomy, University of Maryland, College Park, MD 20742, USA}

\author{Yuki Hirao}%
\altaffiliation{The MOA Collaboration}
\affiliation{\affmoaa}

\author{Andrzej Udalski}
\altaffiliation{The OGLE Collaboration}
\affiliation{Warsaw University Observatory, Al.~Ujazdowskie~4, 00-478~Warszawa, Poland}

\author[0000-0002-2641-9964]{Cheongho Han}
\altaffiliation{The KMTNet Collaboration}
\affiliation{Department of Physics, Chungbuk National University, Cheongju 28644, Republic of Korea}

\author{Ian A. Bond}%
\altaffiliation{The MOA Collaboration}
\affiliation{\affmoaf}

\author[0000-0001-9481-7123]{Jennifer C. Yee}
\altaffiliation{The KMTNet Collaboration}
\affiliation{Harvard-Smithsonian Center for Astrophysics, 60 Garden St., Cambridge, MA 02138, USA}
\nocollaboration

%
%


\author[0000-0003-3316-4012]{Michael D. Albrow}
\affiliation{University of Canterbury, Department of Physics and Astronomy, Private Bag 4800, Christchurch 8020, New Zealand}

\author{Sun-Ju Chung}
\affiliation{Korea Astronomy and Space Science Institute, Daejon 34055, Republic of Korea}
\affiliation{Korea University of Science and Technology, 217 Gajeong-ro, Yuseong-gu, Daejeon 34113, Republic of Korea}

\author{Andrew Gould}
\affiliation{Korea Astronomy and Space Science Institute, Daejon 34055, Republic of Korea}
\affiliation{Department of Astronomy, Ohio State University, 140 W. 18th Ave., Columbus, OH 43210, USA}
\affiliation{Max-Planck-Institute for Astronomy, Königstuhl 17, D-69117 Heidelberg, Germany}

\author[0000-0002-9241-4117]{Kyu-Ha Hwang}
\affiliation{Korea Astronomy and Space Science Institute, Daejon 34055, Republic of Korea}

\author[0000-0002-0314-6000]{Youn-Kil Jung}
\affiliation{Korea Astronomy and Space Science Institute, Daejon 34055, Republic of Korea}

\author[0000-0001-9823-2907]{Yoon-Hyun Ryu}
\affiliation{Korea Astronomy and Space Science Institute, Daejon 34055, Republic of Korea}

\author[0000-0002-4355-9838]{In-Gu Shin}
\affiliation{Harvard-Smithsonian Center for Astrophysics, 60 Garden St., Cambridge, MA 02138, USA}

\author{Yossi Shvartzvald}
\affiliation{IPAC, Mail Code 100-22, Caltech, 1200 E. California Blvd., Pasadena, CA 91125, USA}

\author{Weicheng Zang}
\affiliation{Physics Department and Tsinghua Centre for Astrophysics, Tsinghua University, Beijing 100084, People's Republic of China}

\author{Wei Zhu}
\affiliation{Department of Astronomy, Ohio State University, 140 W. 18th Ave., Columbus, OH 43210, USA}

\author{Sang-Mok Cha}
\affiliation{Korea Astronomy and Space Science Institute, Daejon 34055, Republic of Korea}
\affiliation{School of Space Research, Kyung Hee University, Yongin 17104, Republic of Korea}

\author{Dong-Jin Kim}
\affiliation{Korea Astronomy and Space Science Institute, Daejon 34055, Republic of Korea}

\author{Hyoun-Woo Kim}
\affiliation{Korea Astronomy and Space Science Institute, Daejon 34055, Republic of Korea}

\author{Seung-Lee Kim}
\affiliation{Korea Astronomy and Space Science Institute, Daejon 34055, Republic of Korea}
\affiliation{Korea University of Science and Technology, 217 Gajeong-ro, Yuseong-gu, Daejeon 34113, Republic of Korea}

\author{Chung-Uk Lee}
\affiliation{Korea Astronomy and Space Science Institute, Daejon 34055, Republic of Korea}
\affiliation{Korea University of Science and Technology, 217 Gajeong-ro, Yuseong-gu, Daejeon 34113, Republic of Korea}

\author{Dong-Joo Lee}
\affiliation{Korea Astronomy and Space Science Institute, Daejon 34055, Republic of Korea}

\author{Yong-Seok Lee}
\affiliation{Korea Astronomy and Space Science Institute, Daejon 34055, Republic of Korea}
\affiliation{School of Space Research, Kyung Hee University, Yongin 17104, Republic of Korea}

\author{Byeong-Gon Park}
\affiliation{Korea Astronomy and Space Science Institute, Daejon 34055, Republic of Korea}
\affiliation{Korea University of Science and Technology, 217 Gajeong-ro, Yuseong-gu, Daejeon 34113, Republic of Korea}

\author[0000-0003-1435-3053]{Richard W. Pogge}
\affiliation{Max-Planck-Institute for Astronomy, Königstuhl 17, D-69117 Heidelberg, Germany}
\collaboration{(The KMTNet Collaboration)}

\author{Fumio Abe}
\affiliation{\affmoab}


\author{Richard~K. Barry}
\affiliation{Astrophysics Science Division, NASA/Goddard Space Flight Center, Greenbelt, MD 20771, USA}

\author{Aparna Bhattacharya}
\affiliation{Astrophysics Science Division, NASA/Goddard Space Flight Center, Greenbelt, MD 20771, USA}
\affiliation{Department of Astronomy, University of Maryland, College Park, MD 20742, USA}

\author{Martin Donachie}
\affiliation{\affmoad}


\author{Akihiko Fukui}
\affiliation{\affmoae}
\affiliation{\affmoan}

\author{Yoshitaka Itow}
\affiliation{\affmoab}

\author{Kohei Kawasaki}
\affiliation{\affmoaa}

\author{Iona Kondo}
\affiliation{\affmoaa}

\author{Naoki Koshimoto}
\affiliation{\affmoal}
\affiliation{\affmoam}

\author{Man Cheung Alex Li}
\affiliation{\affmoad}



\author{Yutaka Matsubara}
\affiliation{\affmoab}


\author{Shota Miyazaki}
\affiliation{\affmoaa}

\author{Yasushi Muraki}
\affiliation{\affmoab}

\author{Masayuki Nagakane}
\affiliation{\affmoaa}



\author{Nicholas~J. Rattenbury}
\affiliation{\affmoad}




\author{Haruno Suematsu}
\affiliation{\affmoaa}

\author{Denis~J. Sullivan}
\affiliation{\affmoag}

\author{Takahiro Sumi}%
\affiliation{\affmoaa}

\author{Daisuke Suzuki}
\affiliation{\affmoaj}

\author{Paul~J. Tristram}
\affiliation{\affmoah}


\author{Atsunori Yonehara}
\affiliation{\affmoai}
\collaboration{(The MOA Collaboration)}

\author[0000-0002-9245-6368]{Rados{\l}aw Poleski}
\affiliation{Department of Astronomy, Ohio State University, 140 W. 18th Ave., Columbus, OH 43210, USA}
\affiliation{Warsaw University Observatory, Al.~Ujazdowskie~4, 00-478~Warszawa, Poland}

\author{Przemek Mr{\'o}z}
\affiliation{Warsaw University Observatory, Al.~Ujazdowskie~4, 00-478~Warszawa, Poland}

\author[0000-0002-2335-1730]{Jan Skowron}
\affiliation{Warsaw University Observatory, Al.~Ujazdowskie~4, 00-478~Warszawa, Poland}

\author{Micha{\l} K. Szyma{\'n}ski}
\affiliation{Warsaw University Observatory, Al.~Ujazdowskie~4, 00-478~Warszawa, Poland}

\author{Igor Soszy{\'n}ski}
\affiliation{Warsaw University Observatory, Al.~Ujazdowskie~4, 00-478~Warszawa, Poland}

\author{Szymon Koz{\l}owski}
\affiliation{Warsaw University Observatory, Al.~Ujazdowskie~4, 00-478~Warszawa, Poland}

\author[0000-0002-2339-5899]{Pawe{\l} Pietrukowicz}
\affiliation{Warsaw University Observatory, Al.~Ujazdowskie~4, 00-478~Warszawa, Poland}
\affiliation{Department of Physics, University of Warwick, Gibbet Hill Road, Coventry, CV4~7AL,~UK}

\author{Krzysztof Ulaczyk}
\affiliation{Warsaw University Observatory, Al.~Ujazdowskie~4, 00-478~Warszawa, Poland}
\collaboration{(The OGLE Collaboration)}





\begin{abstract}
  We present the analysis of the microlensing event \ob{15}{1670},
  detected in a high-extinction field, very close to the Galactic
  plane.  Due to the dust extinction along the line of sight, this
  event was too faint to be detected before it reached the peak of
  magnification. The microlensing light-curve models indicate a
  high-magnification event with a maximum of
  $A_\mathrm{max}\gtrsim200$, very sensitive to planetary
  deviations. An anomaly in the light curve has been densely observed
  by the microlensing surveys MOA, KMTNet, and OGLE. From the
  light-curve modeling, we find a planetary anomaly characterized by a
  planet-to-host mass ratio,
  $q=\left(1.00^{+0.18}_{-0.16}\right)\times10^{-4}$, at
  the peak recently identified in the mass-ratio function of
  microlensing planets. Thus, this event is
  interesting to include in future statistical studies about planet
  demography. We have explored the possible degeneracies and find two
  competing planetary models resulting from the $s\leftrightarrow1/s$
  degeneracy. However, because the projected separation is very close
  to $s=1$, the physical implications for the planet for the two
  solutions are quite similar, except for the value of $s$.  By
  combining the light-curve parameters with a Galactic model, we have
  estimated the planet mass $M_2=\MLbval\,\Mearth$ and the lens
  distance $\DL=\DLval\,\kpc$, corresponding to a Neptune-mass planet
  close to the Galactic bulge. Such events with a low absolute
  latitude ($|b|\approx1.1\,\deg$) are subject to both high extinction
  and more uncertain source distances, two factors that may affect the
  mass measurements in the provisional \textit{Wide Field Infrared Survey Telescope} fields. More
  events are needed to investigate the potential trade-off between the
  higher lensing rate and the difficulty in measuring masses in these
  low-latitude fields.
\end{abstract}




\section{Introduction} \label{sec:intro}

Gravitational microlensing has been continuously developed for the
past decades and has proved to be a powerful way to probe the mass
content of our galaxy \citep{Paczynski.1986}. It is a choice method not
only to detect new stellar and substellar objects that are too faint
to be observed otherwise \citep{Mao1991} but also to find stellar
black hole candidates that inhabit the Milky Way \citep{Bennett2002,
  Mao2002, Poindexter2005, Wyrzykowski2016}. Because microlensing does
not rely on the detection of light from the lens, it has a unique
niche among the planet detection techniques for discovering exoplanet
systems at Galactic distances consisting of low-mass planets
\citep{Bennett1996} at large orbital separation \citep{Gould1992}.

To date, more than 3700 confirmed exoplanets, including more than 600 multiple-planets systems have been detected \citep[\eg,][]{Schneider2011}. The
NASA \textit{Kepler} space mission has mostly driven these discoveries
thanks to its unprecedented sensitivity to exoplanets in close orbits
about their host stars \citep{Petigura2013, Burke2015,
  Coughlin.2016}. While transits have become the main exoplanet
detection technique, the radial velocity ground-based surveys have
also contributed substantially to the detection and
characterization of new planets \citep{Bakos2002,
  Pollacco2006}. Despite a large sample of objects that now allow more
robust statistical studies, our understanding of the formation and
evolution of planetary systems remains modest. This is mainly due to
selection effects: most of the exoplanets we know have orbital
separations much smaller than $1~\au$ because of the high sensitivity
of \textit{Kepler} and radial velocity searches to planets at small
separation.

Although the gravitational-microlensing detection technique has found
a modest number of exoplanets up to now (71 planets), these exoplanets
completely dominate the distribution of planets beyond the ``snow
line'' and below 1~Saturn mass. The snow line marks the inner
boundary of the protoplanetary disk where planet formation is 
most efficient, according to the core accretion theory
\citep{Lissauer1987, Lissauer1993, Pollack1996} mostly because ices
can condense in this region \citep{Ida.2004}, which increases the
density of solids by a factor of a few. This can speed up the initial
steps of the planet formation process and, consequently, enable the
formation of gas giants in some planetary systems.

The most recent statistical study \citep{Suzuki.2016} based on the detection of 30
exoplanets by microlensing (the largest sample for such an
investigation until now) found some evidence to support the core
accretion model predictions for planets beyond the snow line. In
particular, this study has discovered a break and a possible peak in
the planet--to--host star mass-ratio function for a mass ratio
$q\approx10^{-4}$. These results have been
supplemented at the low-mass end of the mass-ratio function by an
analysis based on seven planets, and that confirms the ``turnover'' in
the mass function \citep{Udalski2018}, first noted by
\cite{Suzuki.2016}. These results are broadly consistent with the
prediction that ``failed Jupiters'' of $\sim 10\, \Mearth$ should be
more common than gas giants, particularly around the low-mass stars
that dominate the microlensing survey sample.
A peak in the mass-ratio function has recently been found in the
occurrence rate of \textit{Kepler} exoplanets, at a mass ratio
$\approx3\text{-}10$ times smaller than for microlensing exoplanets
\citep{Pascucci2018}. Thus, the most common planets inside the snow
line are less massive than those in wider orbits. This is a strong indication
that the mass-ratio function is a fundamental quantity in planet
formation theory \citep{Suzuki.2016, Pascucci2018, Udalski2018}; this
work also emphasizes the importance of studying and comparing both
regimes.
These state-of-the-art analyses expand previous results
\citep{Gould2010, Sumi2010, Cassan.2012, Shvartzvald.2016.stat}, and they demonstrate again
the ability of microlensing observations to approach the theory of
planetary formation from a different angle while exploring the
exoplanets' demography. These studies also show that the observational
constraints on the mass function of low-mass exoplanets
($\lesssim10\Mearth$) rely on a small number of objects. Meanwhile,
several international collaborations are conducting high-cadence
ground-based surveys and follow-up observations toward the Galactic
bulge (see \Sec{sec:obs}) to detect more microlensing planets and
explore the low-mass end of the exoplanet mass function. In the
future, the \textit{Wide Field Infrared Survey Telescope}
\citep[\textit{WFIRST};][]{Spergel2015,Penny.2019.wfirst} is expected to
observe the densest parts of the Galactic bulge during its
microlensing campaign, where the microlensing event rate is thought to
be highest in the near-infrared (NIR).
Only nine planetary events have been detected in the provisional
\textit{WFIRST} fields, including OGLE-2006-BLG-109Lb,c, the first
Jupiter--Saturn analog found through microlensing
\citep{Gaudi.2008,Bennett2010..ob060109}; MOA-bin-1Lb, a
$3.7\,\Mjup$ super-Jupiter planet \citep{Bennett2012..moabin1};
MOA-2011-BLG-293Lb, the first super-Jupiter in the Galactic bulge and
possibly in the habitable zone detected by microlensing
\citep{Yee2012,Batista2014..293}; OGLE-2013-BLG-0341Lb, a terrestrial
planet in a $1\,\au$ orbit around one member of a $15\,\au$ stellar
binary \citep{Gould.2014.triple}; OGLE-2015-BLG-0966Lb, a cold
Neptune-mass planet in the Galactic disk \citep{Street2016..ob150966};
the Saturn-mass planet OGLE-2013-BLG-1721Lb \citep{Mroz2017a};
OGLE-2013-BLG-1761Lb, a super-Jupiter planet \citep{Hirao2017};
OGLE-2017-BLG-0173Lb, a super-Earth-mass planet
\citep{Hwang2018..ob170173}; KMT-2016-BLG-0212Lb, possibly a
sub-Neptune-mass companion \citep{Hwang2018}; and MOA-2011-BLG-291Lb,
a typical Neptune-mass planet \citep{Bennett.2018.mb110291}.

In this article, we present the analysis of the microlensing event
\ob{15}{1670}, which has two features worthy of special notice.
First, it is in a high-extinction region of the Galactic bulge that is expected to be within the
\textit{WFIRST} footprint. In these fields, the source distance is
more uncertain because the higher stellar density makes more likely
events due to a source lying in the Galactic disk. Excess extinction
and uncertain source distance both may affect the accuracy of the lens
mass measurement. The study of events close to the Galactic plane
similar to \ob{15}{1670} with high-resolution follow-up is of prime
interest to develop the \textit{WFIRST} primary mass measurement method
and characterize the potential trade-off between a higher lensing
rate at low Galactic latitude $\left|b\right|$ (hereafter referred to ``low
$|b|$'') and the difficulty in
determining the masses. Second, this analysis yields the discovery
of a Neptune-mass exoplanet with a mass ratio close to a possible peak
in the mass-ratio function identified in \citet{Suzuki.2016}, where
additional observational constraints are required to strengthen the
statistical results. We present the observations included in the
analysis in \Sec{sec:obs}. \Sec{sec:modeling} describes the
microlensing light-curve modeling. In \Sec{sec:lens}, we use Bayesian
analysis to combine the light-curve models with Galactic priors to
derive an estimate of the planet mass. Finally, we discuss the results
and implications of this work in \Sec{sec:discussions}.

\section{Observations} \label{sec:obs}

The microlensing event \ob{15}{1670} was discovered by the Optical
Gravitational Lensing Experiment \citep[OGLE, phase
IV;][]{Udalski2015} and first alerted on the Early Warning System
(EWS) website on 2015 July 19 at UT 18:34
($\thjd\approx7,223.27$\footnote{$\thjd=\mathrm{HJD}-2,450,000$.}). The
event is located at the J2000 equatorial coordinates
$\left(\mathrm{RA,\ decl.}\right) =
(17^\mathrm{h}\,52^\mathrm{m}\,38\overset{^\mathrm{s}}{.}11,\ 
-28\arcdeg33\arcmin06\overset{\prime\prime}{.}\,9)$, or Galactic coordinates
$(l,\,b)=(1.12105\arcdeg,\,-1.12048\arcdeg)$, in the OGLE-IV field
``BLG500.20,'' which was observed 3--10 times night$^{-1}$. The OGLE survey
toward the Galactic bulge is performed using the $1.3\,\meter$ Warsaw
telescope located at Las Campanas Observatory in Chile. The OGLE
photometry was extracted using OGLE's implementation
\citep{Wozniak2000} of the difference imaging analysis (DIA) technique
\citep{Tomaney1996,Alard1998,Alard2000}. We have calibrated the
resulting OGLE-IV $I$-band photometry \citep{Udalski2015} to the
standard Kron--Cousins $I$ passband and corrected the error bars
following the method described in \cite{Skowron2016}.

Just $42\,\minu$ after OGLE, the Microlensing Observations in
Astrophysics \citep[MOA, phase II;][]{Sumi2003} collaboration
independently found this event (at $\thjd\approx7,223.30$) in the
MOA-II field ``gb5'' and labeled it as \mb{15}{379}. MOA observations
were performed using the $1.8\,\meter$ telescope at the Mount John
University Observatory in New Zealand with a high cadence of
$15\,\minu$ in the wide MOA $R$-band filter. No anomaly alert was sent
because the deviation from a single-lens model occurred the night
before the discovery. On 2015 August 24, the MOA member Yuki Hirao
found the anomaly after modeling the 2015 MOA observations and
immediately identified a possible planetary mass ratio. The MOA
photometry was extracted using MOA's implementation \citep{Bond2001}
of the DIA method.

The Korea Microlensing Telescope Network \cite[KMTNet;][]{Kim2016}
also monitored this event with three $1.6\,\meter$ telescopes located
at the Siding Spring Observatory in Australia (KMTA), the Cerro
Tololo Observatory in Chile (KMTC), and the South Africa
Astronomical Observatory (KMTS).  However, the KMTS data have a large
gap over the anomaly and peak of the event and so are excluded from
the present analysis.
The KMTNet photometry is
derived using the DIA software PySIS \citep{Albrow2009}. The event lies
in the KMTNet field ``BLG02,'' which was observed in 2015
at a cadence of 10 minutes. The event was independently discovered by KMTNet as
KMT-2015-BLG-0186 \citep{Kim2018..eventfinder}.

The final data sets consist of 7609 data points that are used to model
the microlensing light curve. They are summarized in
\Tab{tab:datasets}. All the observations were performed in similar
$I$-band filters, except the wide $R/I$ MOA filter, referred as $R_M$. 

The high-magnification event \ob{15}{1670} has a flux variation of
more than 5.5~mag, which makes the error-bar
estimates on the photometry challenging. For such events, the photometry pipelines
typically underestimate the error bars. Thus, for each data set, we
normalized the error bars on magnitudes, $\sigma$, so that the
$\chi^2$ per degree of freedom, $\chisqred=1$ and the cumulative sum
of $\chi^2$ is approximately linear.  We use the normalization law
\citep{Yee2012}
\begin{equation}\label{eq:renormalization}
  \sigma^\prime_i = k \sqrt{\sigma^2 + e_\mathrm{min}^2}\v
\end{equation}
where $\sigma^\prime$ is the normalized error bar, the constant $k$ is
the rescaling factor, and the constant $e_\mathrm{{min}}$ mostly
modifies the highly magnified data. The normalization constants are
given in \Tab{tab:datasets}.

\begin{table}[tbp]{\centering
\caption{Telescopes and Photometric Data Sets}\label{tab:datasets}
\setlength{\tabcolsep}{12.96pt}
\begin{tabular}{@{} l c c c c c @{}}\toprule
    Telescope & Location & Filter & Data\textsuperscript{a} & k\textsuperscript{b} & $e_\mathrm{min}$\textsuperscript{b}\\ \midrule
MOA ($1.8\,\meter$)  & Mount John, New Zealand  & $R_M$\textsuperscript{c} & 4395 & 1.207 & 0.003 \\
KMTC ($1.6\,\meter$) & Cerro Tololo, Chile      & $I$                      & 1032 & 0.499 & 0.003 \\
KMTA ($1.6\,\meter$) & Siding Spring, Australia & $I$                      & 833  & 1.200 & 0.003 \\
OGLE ($1.3\,\meter$) & Las Campanas, Chile      & $I$                      & 821  & 1.381 & 0.003 \\\bottomrule\addlinespace[2ex]
\end{tabular}\par}

{\footnotesize {\bfseries Notes.}

\smallskip\textsuperscript{a}\,Number of observations after data cleaning.

  \smallskip\textsuperscript{b}\,Error-bar rescaling factor.

 \smallskip\textsuperscript{b}\,MOA wide filter corresponding to a Cousins $R$ and $I$ band.}
\end{table}

\section{Light-curve Models} \label{sec:modeling}

\subsection{Lens Parameters} \label{sec:modeling:params}

\begin{figure}[tbp]
  \begin{center}
    \includegraphics[scale=1]{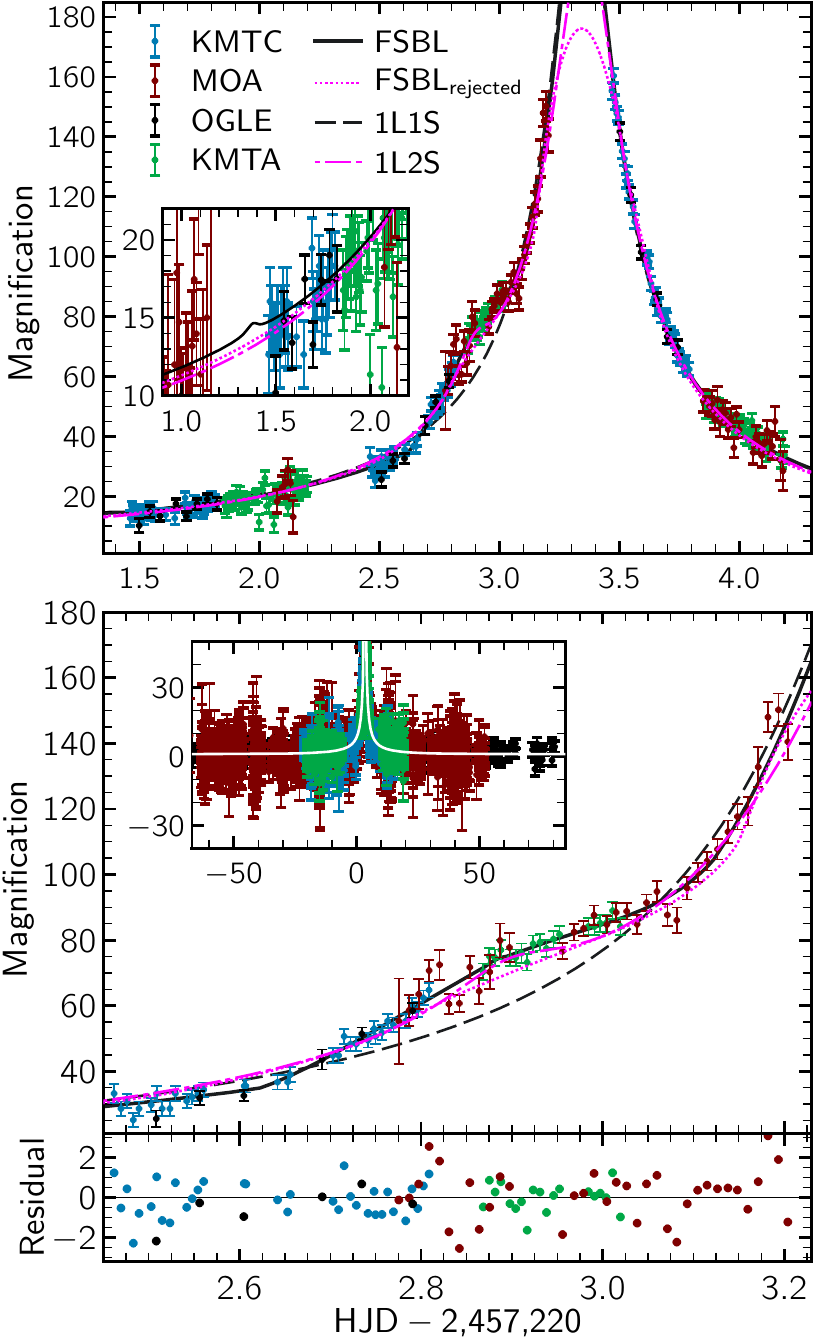}
    \caption{Light curve of the microlensing event \ob{15}{1670} and
      the best binary lens with a finite source effects model (FSBL;
      solid line). For comparison, the dotted line shows the rejected
      model with a $q\approx10^{-3}$ mass ratio
      ($\mathrm{FSBL_{rejected}}$), the dashed line shows the
      best-fit single-lens model (1L1S), and the
      dashed-dotted line (1L2S) refers to the single-lens binary-source
      model (see \Sec{sec:modeling:params}). Each color refers to one
      observatory (MOA in red, KMTC in blue, KMTA in green, and OGLE in
      black). In the lower panel, the residuals are plotted in
      $\sigma$ units, and the inset shows the full light curve in a
      time ($\mathrm{HJD}-2,457,220$) vs. magnification plot along
      with the best-fit model in white. In the upper panel, the inset
      shows the magnification during the cusp approach (see
      \Fig{fig:caustics}).}
    \label{fig:lightcurve}
  \end{center}
\end{figure}

The light curve of this event, shown in \Fig{fig:lightcurve}, looks
very much like a single-lens event, except during the short time
interval $\thjd\in [7,222.6,\,7,223.1]$, close to the peak of
magnification. In this interval, the observations of the four
observatories (MOA, KMTC, KMTA, and OGLE) caught a clear bump (the anomaly) in
the light curve corresponding to a deviation from a single-lens
model. This deviation typically occurs when the ``major image''
created by a host star is perturbed by the gravity of a companion,
possibly a planet. This image moves in the vicinity of the lens
Einstein ring during the lens-source relative motion, at an angular
separation from the host star close to the angular Einstein radius,
\begin{equation}\label{eq:thE}
  \thE = \sqrt{\frac{4G\ML}{c^2\DS}\left(\frac{\DS}{\DL}-1\right)}\v
\end{equation}
where $G$ is the gravitational constant, $c$ is the speed of light,
$M$ is the total mass of the lens, and $\DL$ and $\DS$ are,
respectively, the observer-lens and observer-source distances.
Consequently, such a perturbation is very likely when the companion is
located close to the Einstein ring of the host star \citep{Griest.1998}. The single-lens
model indicates a high-magnification event that is very sensitive to the
detection of planets around the peak of magnification, \ie, when the
multiple images created by the host are very much elongated around the
Einstein ring. Hence, in this context, the anomaly is compatible with a source star
that crosses a caustic.

We start modeling the light curve based on a point-source single-lens model (hereafter ``1L1S'') that does not require any large computing power while
providing a first estimate of the most fundamental parameters. During
this process, we fit the event with a Paczy\'nski light curve
\citep{Paczynski.1986} that depends on three parameters: the impact
parameter of the apparent source trajectory relative to the lens,
$u_0$; the time at which the source reaches $u_0$, $t_0$; and the
Einstein radius crossing time, $\tE = \thE/\murel$, where $\murel$ is
the lens-source relative proper motion.

Three additional parameters are required to model a binary lens: the
mass ratio of the secondary to primary lens component
$q = \MLb/\MLa$, where $\MLb$ ($\MLa$) is
the mass of the secondary lens (the mass of the primary
lens, with $\ML=\MLa+\MLb$); the separation in Einstein units, $s$;
and the angle between the lens axis and the source trajectory,
$\alpha$. For a binary lens, $u_0$ is the distance of closest approach
between the lens center of mass and the source.
Due to the possibility that the lens crosses or approaches close to a caustic,
we take into account
the physical size of the source, \ie, the finite source effects, by
adding one model parameter, namely, the source radius crossing time,
$\tS=\rho\,\tE=\thS/\murel$, where $\rho$ is the source angular radius
in Einstein units, \ie,
\begin{equation}\label{eq:rho}
  \rho=\frac{\thS}{\thE}\v
\end{equation}
with $\thS$ the source angular radius. The source crossing time links
the parameters used in the fit and two fundamental physical
quantities: the angular Einstein radius and the lens-source relative
proper motion. Hereafter, we refer to the resulting ``finite-source
binary-lens'' model as ``FSBL.''

Finite source effects in microlensing light curves are usually
sensitive to the stellar limb darkening \citep{Albrow1999,
  Cassan2006}. We include this effect in the model by considering a
source described as a nonuniform disk \citep{An2002, Zub2011} with
the linear intensity-normalized profile
\begin{equation}\label{eq:limb}
  I(r) = \frac{1}{\pi}\left[ 1 - \Gamma \left( 1 - \frac{3}{2} \sqrt{1 - r^2} \right) \right]\v
\end{equation}
where $\Gamma$ is a linear limb-darkening coefficient and $r$ is the
fractional distance from the center toward the limb of the star (\ie,
$r\in[0,1]$). The linear \eq{eq:limb} is generally a good
approximation, in particular when the limb darkening is weakly
constrained, \eg, for a particularly faint event like \ob{15}{1670}.
We use the extinction-free source color found in \Sec{sec:rho} to
estimate the effective temperature of the source,
$T_\mathrm{eff}\approx 4600\,\mathrm{K}$, and its surface gravity,
$\log g\approx 4.5$. For these values and adopting a metallicity
$\log [M/H] = 0$, we adopt the linear limb-darkening coefficients
$u_I=0.6155$ (\ie, $\Gamma_I=0.5163$) and $u_R = 0.7259$ \citep[\ie,
$\Gamma_R=0.6384$;][]{Claret2011}.

Finally, two parameters describe the unlensed source flux:
$\fsl{j,\lambda_i}$, for any observatory, $j$, and passband, $\lambda_i$, and the excess flux,
$\fbl{j,\lambda_i}$, resulting from the combination of any (and possibly
several) ``blend'' stars.
The blend can be either the lens itself or an unrelated star or stars.
At any time $t$, the total flux of the
microlensing target is
\begin{equation}\label{eq:ftotal}
  F_{j,\lambda_i} (t) = A(t) \fsl{j,\lambda_i} + \fbl{j,\lambda_i}\v
\end{equation}
where $A(t)$ is the source flux magnification at the date $t$.  During
the fitting process, for each set of nonlinear fit parameters and
each passband, we solve the linear \eq{eq:ftotal} \citep{Rhie1999}. In
practice, $\lambda_i$ is the $I$ and $R$ filters.  The source
magnitude reported in \Tab{tab:model_parameters}, $\Is$, is derived
after the OGLE-IV photometry calibration.

\subsection{Exploration of Parameters Space} \label{sec:exploration}

\subsubsection{Single-source Binary-lens Model}\label{sec:1s2l}

The best-fit 1L1S model is used as a starting point to explore
binary-lens models. Computing the source flux magnification for a
high-magnification event is usually time-consuming. Several numerical
methods have been developed to optimize the computational cost, such
as image contouring methods \citep{Gould1997,Dominik2007,Bozza2010}
or ray-shooting techniques \citep{Bennett1996,Dong2006,Dong2009}.
During the light-curve modeling process, we use the image-centered
ray-shooting method \citep{Bennett1996}. We start exploring possible
FSBL solutions using the initial condition grid search method
described in \citet{Bennett.2010} for
$\log\left(0.3\right) \leqslant \log{s} \leqslant \log\left(12.5\right)$ and
$-4\leqslant \log{q} \leqslant -0.954$. The three parameters
$\{s,\, q,\, \alpha\}$ are fixed, while the other parameters
vary. We use a Monte Carlo approach to perform a global search using
a Metropolis algorithm with an adaptive size of the proposal
function to find the best-fit models.  For each model, we compute the
$\chi^2$ value. The local minima of the $\chi^2$ function correspond
to plausible physical models; we select the solutions with
$\Delta\chi^2 = \chi^2-\chi^2_\mathrm{min} \leqslant 150$ for a refined
exploration that allows all parameters to vary during a Markov chain
Monte Carlo (MCMC) to sample the posterior probability distribution.

\begin{table}[tbp]
{\centering
\caption{Parameters for the Best-fit Model and the
    Corresponding Statistical Values from the Posterior Probability
    Distribution Function \label{tab:model_parameters}}\label{tab:observ}
{\footnotesize
\begin{tabular}{@{} l @{\hspace*{6pt}}l @{\hspace*{8pt}} r r r c c @{}}\toprule
& & \multicolumn{3}{c}{Best Fit} & \multicolumn{2}{c}{MCMC (95.5\% Confidence Interval)}\\ \cmidrule(lr){3-5} \cmidrule(l){6-7}
Parameter & Units & $s<1$ & $s>1$ & $s>1$ & $s<1$ & $s>1^\text{a}$\\ \midrule
$\chi^2$         &         & 7052.8\phnfo       & 7042.8\phnfo       & 7046.2\phnfo       & \nodata                   & \nodata\\
$\Delta\chi^2$   & \nodata & \phntw10.0\phnfo   & \phnth0.0\phnfo    & \phnth3.4\phnfo    & \nodata                   & \nodata\\
$q/10^{-4}$      & \nodata & \phnth1.12809      & \phnth0.79836      & \phnth0.89779      & $1.50^{+0.86}_{-0.68}$    & $1.00^{+0.40}_{-0.31}$\\
$s$              & \nodata & \phnth0.96318      & \phnth1.03529      & \phnth1.05331      & $0.965\pm 0.010$          & $1.056^{+0.028}_{-0.020}$\\
$\tE$            & days    & \phntw35.19112     & \phntw27.91693     & \phntw23.94770     & $27.0^{+12}_{-7.5}$       & $23.3^{+9.1}_{-5.2}$\\
$\tS/10^{-2}$    & days    & \phnth6.08626      & \phnth3.19663      & \phnth5.05031      & $6.0\pm1.1$               & $5.05^{+0.60}_{-1.7}$\\
$t_0$            & $\thjd$ & 7223.34497         & 7223.34246         & 7223.34247         & $7223.3454\pm 0.0036$     & $7223.3427\pm 0.0033$\\
$u_0/10^{-3}$    & \nodata & \phnth3.52282      & \phnth4.30742      & \phnth5.08333      & $4.7^{+1.9}_{-1.5}$       & $5.3^{+1.7}_{-1.5}$\\
$\alpha$         & rad     & \phnth0.26894      & \phnth0.25528      & \phnth0.25522      & $0.275^{+0.020}_{-0.024}$ & $0.257^{+0.015}_{-0.013}$\\ \midrule
$\rho/10^{-3}$   & \nodata & \phnth1.72949      & \phnth1.14505      & \phnth2.10889      & $2.23^{+0.86}_{-0.71}$    & $2.17^{+0.69}_{-1.0}$\\
$\Is$            & \nodata & \phntw22.809\phntw & \phntw22.540\phntw & \phntw22.371\phntw & $22.51^{+0.40}_{-0.37}$   & $22.34^{+0.40}_{-0.30}$\\ \bottomrule \addlinespace[6pt]
\end{tabular}} \par}

{\footnotesize {\bfseries Notes.} The uncertainties correspond to a 95.5\% confidence interval, and the measurement is the median of the posterior. The parameter $\rho=\tS/\tE$ is not fit.

\smallskip \textsuperscript{a}\,We include in this column the two degenerate solutions with $s>1$ because their two respective non-Gaussian posterior distributions are connected. As the volume of the parameter space that corresponds to a given confidence level is much larger in the vicinity of the solution with $s \approx 1.05$, the overall posterior probability close to that solution is higher. See discussion in \Sec{sec:1s2l}.}
\end{table}

\begin{figure}[tbp]
{\centering
  \epsscale{1} \includegraphics[scale=1]{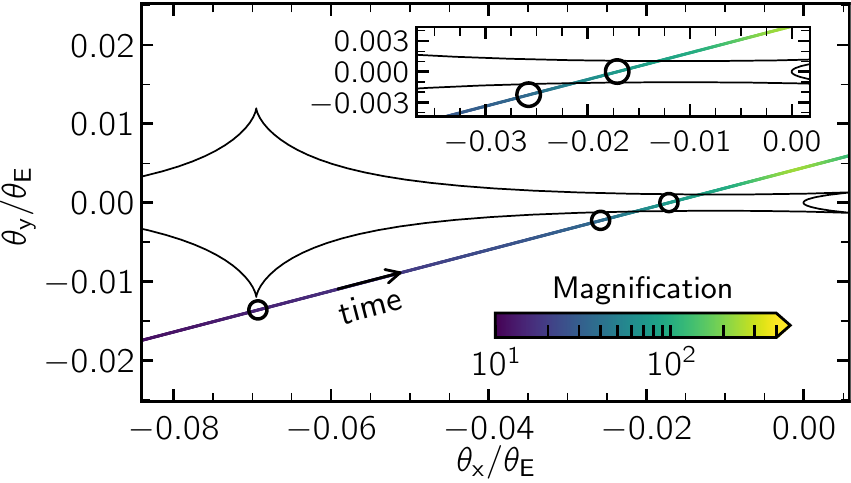}\\
  \medskip \includegraphics[scale=1]{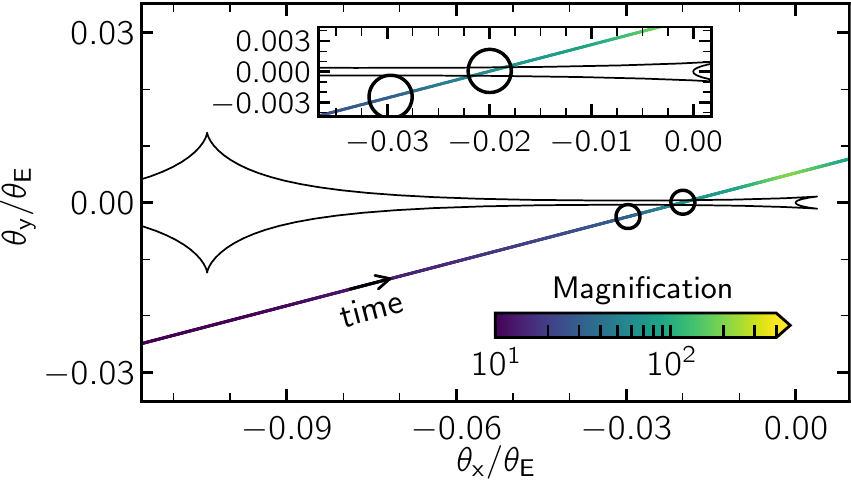}\\
  \medskip \includegraphics[scale=1]{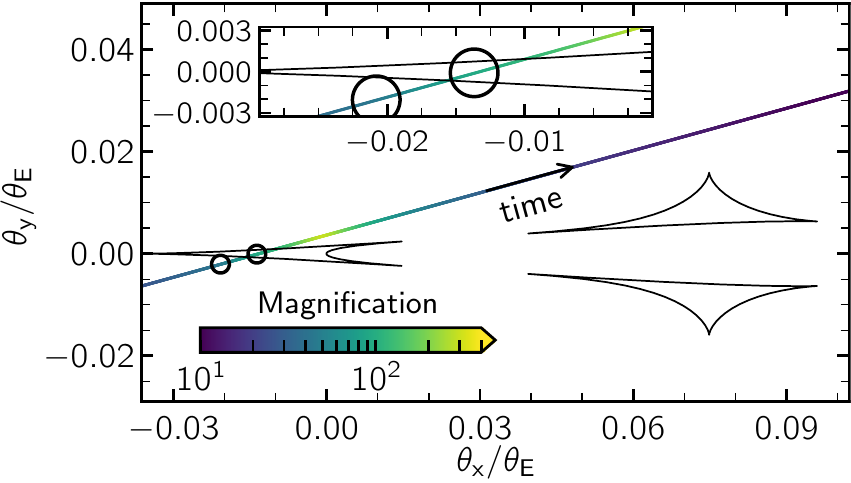}\par}
  \caption{Caustic topology of the two best-fit models ($s=1.03529$,
    top panel; $s=1.05331$, middle panel) and the corresponding
    degenerate solution ($s<1$, bottom panel), shown by the black
    line. The colored line refers to the source trajectory relative to
    the lens, and the inset shows a zoom-in on the caustic crossing.
    In the top panel, the source edge is drawn at $\thjd = 7,221.375$,
    the time of the caustic entry ($\thjd = 7,222.63$), and
    $\thjd = 7,222.88$ (time of the peak of the planetary anomaly). In
    the middle and bottom panels, it is drawn at the time of the caustic
    entry ($\thjd = 7,222.64$ and $7,222.62$, respectively) and
    $\thjd=7,222.88$ (same as the top).  The color along the source
    trajectory refers to the magnification, and the arrow shows the
    direction of the source-lens relative motion. The caustic is shown
    in the center-of-mass reference frame, with the planet (the host
    star) on the left-hand side (right-hand side).}
  \label{fig:caustics}
\end{figure}

The best-fit models for this event have a planetary mass ratio. These
planetary solutions are favored over a single-lens model by
$\Delta\chi^2=932$. In particular, two main models (and their
degenerate solutions) were identified during the refined exploration
of the parameter space: one with $q=1.19\times10^{-3}$, which is
ruled out by $\Delta\chi^2\approx109$, compared to the best-fit model
with $q=7.98357\times10^{-5}$.  The best-fit model parameters are
presented in \Tab{tab:model_parameters} and the model light curves are
plotted in \Fig{fig:lightcurve} (hereafter the ``FSBL model'').  As
we can see in this figure, the best-fit model provides a better
explanation for both the caustic entry and the anomaly than the
higher mass-ratio solution (hereafter the
``$\mathrm{FSBL_{rejected}}$ model''). Note that we have chosen the
FSBL model as a reference in \Fig{fig:lightcurve}; \ie, we plot the
corrected magnification,
\begin{equation}
  A_{i,\mathrm{plot}}(t) = \frac{f_{\mathrm{s},i}}{f_{\mathrm{s,ref}}} A_i(t) + \frac{f_{\mathrm{b},i} - f_{\mathrm{b,ref}}}{f_{\mathrm{s,ref}}}\v
\end{equation}
where $A_i$ is the magnification derived for the model
$i=\{\mathrm{FSBL_{rejected}, 1L1S, 1L2S}\}$, $f_{\mathrm{s},i}$ and
$f_{\mathrm{b},i}$ are the source and blend flux for the model $i$,
and $f_{\mathrm{s,ref}}$ and $f_{\mathrm{b,ref}}$ are the calibrated
source and blend flux derived from the reference model.  The best-fit
model describes an intermediate binary configuration (resonant caustic
with $s=1.03529$ and $q=7.98357\times10^{-5}$) shown in the top
panel of \Fig{fig:caustics}. The source trajectory passes close to the
host star, responsible for the high-magnification values. Also, the
caustic crossing happened in one of the thinnest regions of the
caustic (slightly thinner than the source size), resulting in a
moderate deviation from a single-lens model as shown in
\Fig{fig:lightcurve}. The magnification derived from the best-fit
model reaches $A_\mathrm{max}\approx232$. This solution also includes
a cusp approach before the source crosses the caustic and during a gap
in the observations at $\thjd\approx7,221.4$ (see upper inset in
\Fig{fig:lightcurve}).

This best-fit model is degenerate with another slightly different
solution characterized by $s=1.05331$ and $q=8.97794\times10^{-5}$,
disfavored by only $\Delta\chi^2=3.4$. As shown in
\Tab{tab:model_parameters}, this solution has a higher source crossing
time and slightly shorter Einstein timescale, resulting in a source
radius approximately twice as large as the value derived from the best-fit model ($\rho=2.1\times10^{-3}$ versus $1.1\times10^{-3}$). The
caustic topology and source trajectory are similar to the best-fit
model and are shown in the middle panel of \Fig{fig:caustics}. The source
crosses a resonant caustic in a region where the distance between the
two caustic edges is smaller than the source radius. The degeneracy between
the two solutions with $s>1$ is due to a degeneracy between the source size and
the width of the caustic that occurs when the anomaly
consists of a smooth ``bump.''

\begin{figure}[tp]
  \epsscale{1}
  \begin{center}
    \includegraphics[width=\linewidth]{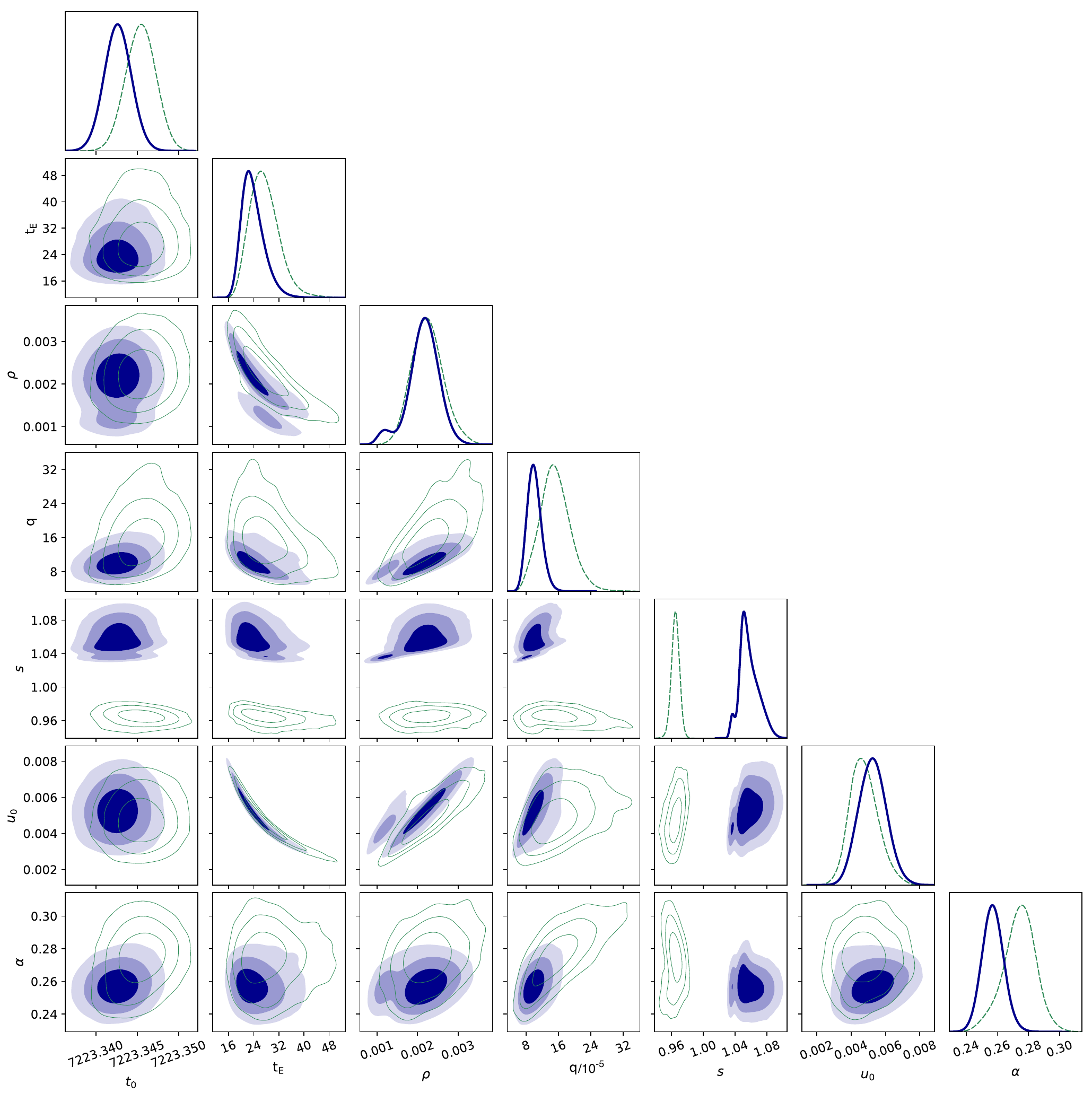}
  \end{center}
  \caption{Correlation between the parameters for the best-fit model
    ($s>1$) in blue and its degenerate alternative ($s<1$) in green
    (see \Sec{sec:exploration}). For each solution, the three shaded
    areas (or contours) show the $68.3\%$, $95.5\%$, and $99.7\%$
    confidence regions, respectively, from the darkest to lightest
    color. The two solutions with $s>1$ are included in the blue
    shaded regions. The units are defined in
    \Tab{tab:model_parameters}.}
  \label{fig:correlations}
\end{figure}

These caustic crossing features can be approximately reproduced by a
close binary-lens configuration with $s=0.96318$ and
$q=1.128086\times10^{-4}$, as shown in the bottom panel of
\Fig{fig:caustics}. This solution corresponds to the well-known
$s\leftrightarrow 1/s$ degeneracy \citep{Griest.1998,Dominik1999},
which is common when a caustic crossing involves the central caustic
in a close binary-lens configuration. For a planetary mass ratio, the
closer $s$ is to~1, the weaker the degeneracy. As the lens parameters are
very close to $s=1$, it is possible to choose between the $s<1$ and
$s>1$ solutions: the latter is favored by $\Delta\chi^2=10.0$. For
$s<1$, we do not find two likelihood maxima. Conversely, all of the MCMC
chains converged to the same solution shown in
\Tab{tab:model_parameters} and characterized by a source size of
$\rho=1.73\times10^{-3}$. The two $s<1$ and $s>1$ degenerate solutions
are very close in terms of goodness of fit, and the marginal
distributions derived at the end of the MCMC are very much overlapping
for all of the parameters except the separation, $s$. The parameter
correlation and marginal distributions for both solutions are
shown in \Fig{fig:correlations}.  The solution corresponding to
$s=1.05331$ is a local maximum of the likelihood, \ie, one mode of the
posterior distribution. The $\Delta\chi^2$ between these two solutions
with $s>1$ corresponds to a relative probability of $0.18$. However,
because the volume of the parameter space that corresponds to a given
confidence level is much larger in the vicinity of the solution with
$s\approx1.05$, the overall posterior probability close to that
solution is higher. For the next stages of the analysis, we use the
full multimodal posterior to estimate the lens mass and distance in
\Sec{sec:bayes}, including the solution with $s<1$, rather than
selecting the best-fit model only.

\begin{figure}[tbp]{\centering
  \epsscale{1} \includegraphics[scale=1]{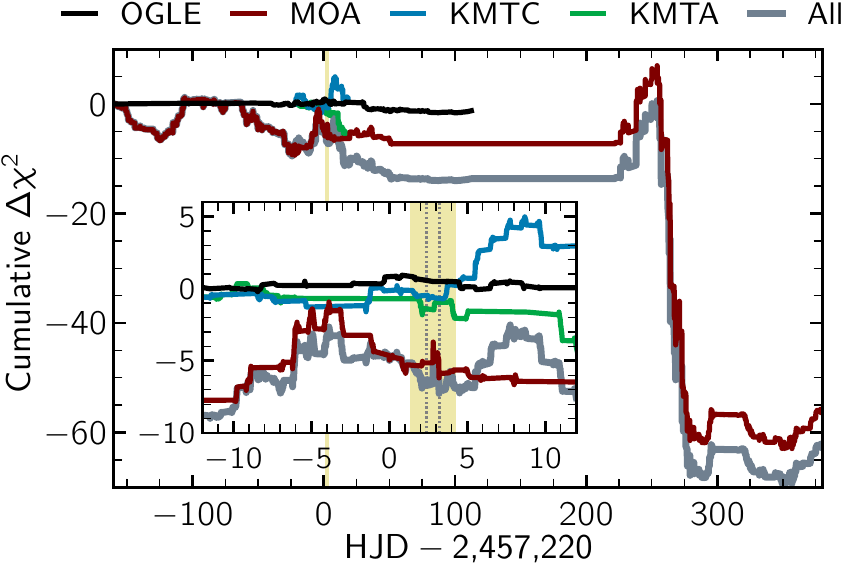}\\ \medskip
  \medskip \includegraphics[scale=1]{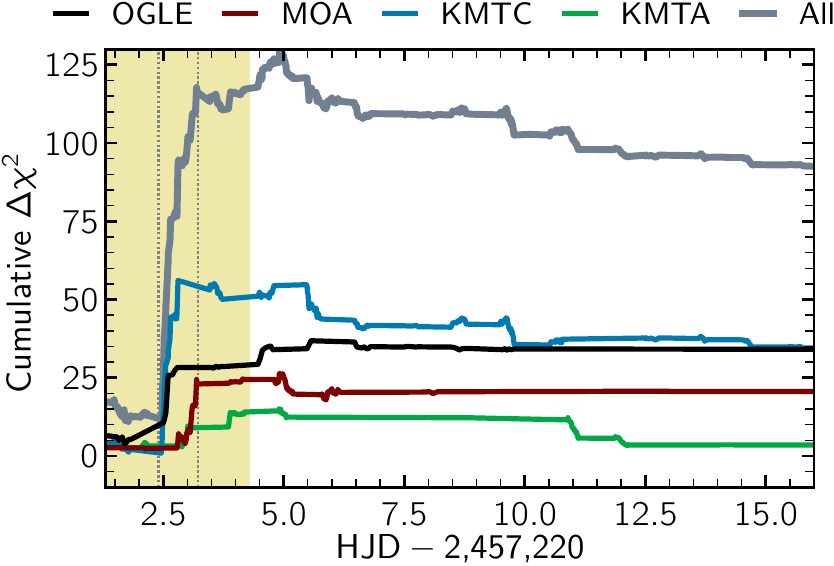}\par}
  \caption{Cumulative $\Delta\chi^2$ for the single-source,
    binary-lens model with parallax (upper panel) and the
    binary-source, single-lens model (lower panel). The best-fit
    planetary model in \Tab{tab:model_parameters} is used as a reference
    in both cases.  The yellow shaded region corresponds to the time
    interval shown in the upper panel of \Fig{fig:lightcurve}, and the
    dotted lines indicate the region where the magnification is
    substantially different than that of a single-source, single-lens model
    ($7222.4<\thjd<7223.22$).}
  \label{fig:cumul}
\end{figure}

We also searched for a possible parallax detection in the light curve.
During this event, the Earth's instantaneous acceleration in the
heliocentric reference frame projected to the lens plane was only
$\approx 50\%$ of its maximum. Indeed, the peak of magnification was
reached on 2015 July 19, less than a month after the minimum of the
Earth's acceleration perpendicular to the line of sight. Additionally,
this event is faint, and the uncertainties
make it more difficult to detect
asymmetric features in the light-curve tails.
The best-fit model with parallax is favored over the static model by
$\Delta \chi^2 = 54$. This model has a secondary magnification peak
during the gap between the 2015 and 2016 observing seasons. The upper
panel of \Fig{fig:cumul} shows the cumulative $\Delta\chi^2$ between
the model including parallax compared to the best-fit static
solution.  As we can see in this figure, the overall
$\chi^2$ improvement mostly comes from baseline observations performed
by MOA during the 2016 observing season ($7470 \leqslant \thjd \leqslant 7500$),
likely due to fluctuation in the baseline data. Out of the overall
$\chi^2$ improvement of $54$, there is an improvement of only
$\Delta\chi^2 \approx 6$ for $\thjd \leqslant 7231$, mostly due to data
points from MOA and KMTA: the improvement is, respectively,
$\Delta\chi^2\approx4.4$ and $1.7$ for observations
when the magnification is $A\geq3.5$ (the noise in magnification is
typically $\pm2.5$).  Meanwhile, numerous data points from KMTC favor
the static model by $\Delta\chi^2\approx 5$ during the same time
interval. We conclude that the overall improvement when the
magnification emerges from the noise in the baseline is
$\Delta\chi^2\leqslant0.5$. In summary, $90\%$ of $\chi^2$ improvement for
the model with parallax comes from baseline data, when the
magnification is $A \leqslant 1.03$, and the remaining $10\%$ is due to
data points at low magnification and brightness (the target
brightness is $I\approx19.3$ when $A=10$). For these reasons, we do
not claim a parallax detection in the light curve of this event.
As a consequence, an absolute mass measurement of the lens
\ob{15}{1670L} components will not be possible with the light-curve
data alone, but the high precision
on the planet-to-host mass ratio will be enough to identify the
physical nature of the planetary component (see \Sec{sec:lens}).

\subsubsection{Binary-source Single-lens Model}\label{sec:2s1l}

\begin{table}[htbp]
{\centering
\caption{Parameters for the best-fit 1L2S Including the Source Orbital Motion}\label{tab:2s1l_params}
\setlength{\tabcolsep}{12pt}
\begin{tabular}{@{} l c r @{}}\toprule
Parameter & Units & \multicolumn{1}{c}{Value}\\ \midrule
$\chi^2$            & \nodata            & 7137.4\phnfo     \\
$\Delta\chi^2$      & \nodata            & \phntw94.6\phnfo \\
$\tE$               & days               & \phnth6.73528    \\
$t_0$               & $\thjd$            & 7223.35787       \\
$u_0/10^{-3}$       & \nodata            & \phntw20.09699   \\
$t_{0,2}$           & $\thjd$            & 7222.89358       \\
$u_{0,2}/10^{-3}$   & \nodata            & \phnth8.21605    \\
$f_{s_2,I}/10^{-2}$ & \nodata            & \phnth3.71410    \\
$f_{s_2,R}/10^{-2}$ & \nodata            & \phnth3.95530    \\
$dt_{\mathrm{E},2}$ & days               & \phnth1.80657    \\
$10^{-2}/T_\mathrm{Sorb}$ & days$^{-1}$  & \phnth6.60615    \\ \bottomrule
\end{tabular}\par}
\end{table}

In the previous section, we have described the modeling strategy we
followed to find the binary-lens model that best fits the light
curve.  For completeness, we also consider possible binary-source,
single-lens models (hereafter called 1L2S), starting with a grid
search method for the source projected separation in Einstein units,
$s_\mathrm{source}$ (300 points for
$10^{-3}\leqslant s_\mathrm{source} \leqslant 1$, and 100 points for
$1\leqslant s_\mathrm{source}\leqslant 5$). To explore 1L2S models, we use the
binary-source, binary-lens modeling code written to model microlensing
event MOA-2010-BLG-117 \citep{Bennett2018..mb100117}. This code uses the
single-lens parameters $t_{0,i}$, $u_{0,i}$, and $t_{\mathrm{E},i}$,
corresponding to the microlensing of the stellar binary component
$i=\{0,1\}$.

To include the orbital motion of the binary source, we introduce
$\diff{}{\tE}= t_{\mathrm{E},2}-t_{\mathrm{E},1}$ to account for the
different lens-source relative motions due to this source's orbital
motion in the direction parallel to the source-lens relative
motion. The orbital motion perpendicular to the source motion can be
described by the difference in the angles that the source-lens
relative motion subtends with respect to the lens system,
$d\theta$. However, because of the circular symmetry of a single-lens
system, neither these angles nor their difference is measurable.
However, when allowing for a circular orbit with period $T_\mathrm{Sorb}$,
as in \cite{Bennett2018..mb100117}, we do need $d\theta$ to describe
the instantaneous velocity of the two sources, although the angle,
$\theta$, remains unmeasurable for a single-lens system.  We use
$1/T_\mathrm{Sorb}$ as our parameter to describe the orbital period. The
reference time when the sources are at their reference positions and
velocities is $\thjd=7223.335$.  We use the parameters
$t_{\mathrm{E},1}$ and $dt_{\mathrm{E}}$ instead of the
two independent Einstein timescales.  In order to avoid unphysical
regions of the parameter space, we impose the condition that the
source 2--to--source 1 flux ratio must be the same for all data sets
taken in the same passband. Thus, we fit two parameters, $f_{s2,I}$
and $f_{s2,R}$, one for each filter used to obtain the data.
We
have explored the parameter space using an MCMC algorithm, and we find
that the best binary-lens model is favored over the best binary-source
model by $\Delta\chi^2=95$. The best-fit 1L2S model parameters are
shown in \Tab{tab:2s1l_params}, and the lower panel of \Fig{fig:cumul}
is the cumulative $\Delta\chi^2$ between the 1L2S
binary-source model. \Fig{fig:cumul}
indicates that the binary-lens model is highly favored by
$\Delta\chi^2\approx 95$. In
particular, $\Delta\chi^2\approx115$ arises from a time window
corresponding to the anomaly. As a consequence, a 1L2S model does not
compete with the binary-lens alternatives presented in
\Tab{tab:model_parameters}.

\subsubsection{Robustness of Best-fit Solutions}

The error-bar normalization law adopted in \eq{eq:renormalization} might be sensible and is standard practice when dealing with DIA photometry.
In the case of OGLE-2015-BLG-1670, we investigate the effect of a moderate change of $e_\mathrm{min}$ on the robustness of the best-fit solutions reported in \Tab{tab:model_parameters}. We consider the five following situations:
\begin{enumerate}
    \item $e_\mathrm{min}(\text{all}) = 0$;
    \item $e_\mathrm{min}(\text{KMTA, KMTC}) = 0$, $e_\mathrm{min}(\text{others}) = 0.003$;
    \item $e_\mathrm{min}(\text{KMTA}) = 0$, $e_\mathrm{min}(\text{others}) = 0.003$;
    \item $e_\mathrm{min}(\text{KMTC}) = 0$, $e_\mathrm{min} = 0.003$ otherwise; and
    \item $e_\mathrm{min}(\text{KMTA}) = 0$, $e_\mathrm{min}(\text{KMTC}) = 0.006$, $e_\mathrm{min}(\text{others}) = 0.003$.
\end{enumerate}
In each case, we compute $k$ so that $\chisqred=1$, and we run several MCMCs from the plausible physical models identified after the grid search, following the same method as described in \Sec{sec:exploration}. The resulting best-fit models are very close to the ones originally identified. Then, we run refined MCMCs from these models. At this stage, we find solutions within $1\sigma$--$2\sigma$ of the model parameters reported in \Tab{tab:model_parameters} and~\ref{tab:2s1l_params}. Finally, we run a last set of MCMCs using the parameters from \Tab{tab:model_parameters} and~\ref{tab:2s1l_params} as initial conditions.

As expected, we find a slightly different $\chi^2$ difference between each model. However, the best-fit parameters do not change, and this very limited change in the $\chi^2$ difference shows that the conclusions do not depend on the fine details of the coefficients used in the error-bar normalization law assumed. First, the high mass ratio model ($q\sim1.2\times 10^{-3}$) remains disfavored by a $\chi^2$ difference greater than 100 in all cases. Second, the 2L1S low mass ratio solution is highly favored by $\Delta\chi^2 > 96$ compared to the 1L2S model in all cases. Finally, regarding planetary solutions, the most significant difference is obtained in case 4: $\chi^2(s=1.05)-\chi^2(s=1.03) = 2.36$, and $\chi^2(s<1)-\chi^2(s=1.03) = 9.10$. The change in the $\chi^2$ difference remains very small, and in that particular case, KMTC error bars are assumed to be smaller than they should be. Conversely, the less significant changes are found in case 5, where $e_\mathrm{min}$ is assumed to be twice the value originally chosen.

We conclude that the best-fit model parameters are robustly determined, and the results are not too sensitive to a moderate change of the error-bar  normalization coefficients.
The overlap of observations from multiple surveys partly explains this robustness. Indeed, the data from the four surveys cover the anomaly but also many portions of the light curve. For instance, OGLE, MOA, KMTC, and KMTA observed during the event, at the anomaly, and at the baseline. Similarly, OGLE, MOA, and KMTC have numerous simultaneous observations when the magnification is high.

\section{Lens Physical Properties} \label{sec:lens}

\subsection{Measurement of the Angular Einstein Radius}\label{sec:rho}

The measurement of the angular Einstein radius provides one relation
between the lens mass and distance. Indeed, from
equation~(\ref{eq:thE}), the lens total mass reads
\begin{equation}\label{eq:mass_from_rho}
  \ML = \frac{c^2\,\thE^2}{4\G} \left(\frac{1}{\DL}-\frac{1}{\DS}\right)^{-1}\p
\end{equation}
Modeling the microlensing light curve yields a precise measurement of
$\rho$, as well as the source flux. By combining the latter
quantity with a color-magnitude diagram (CMD) of stars from the same
field of view as the target, it is possible to measure the source
color and determine its angular radius, $\thS$.

\begin{figure}[tbp]
  \begin{center}
    \includegraphics[scale=1]{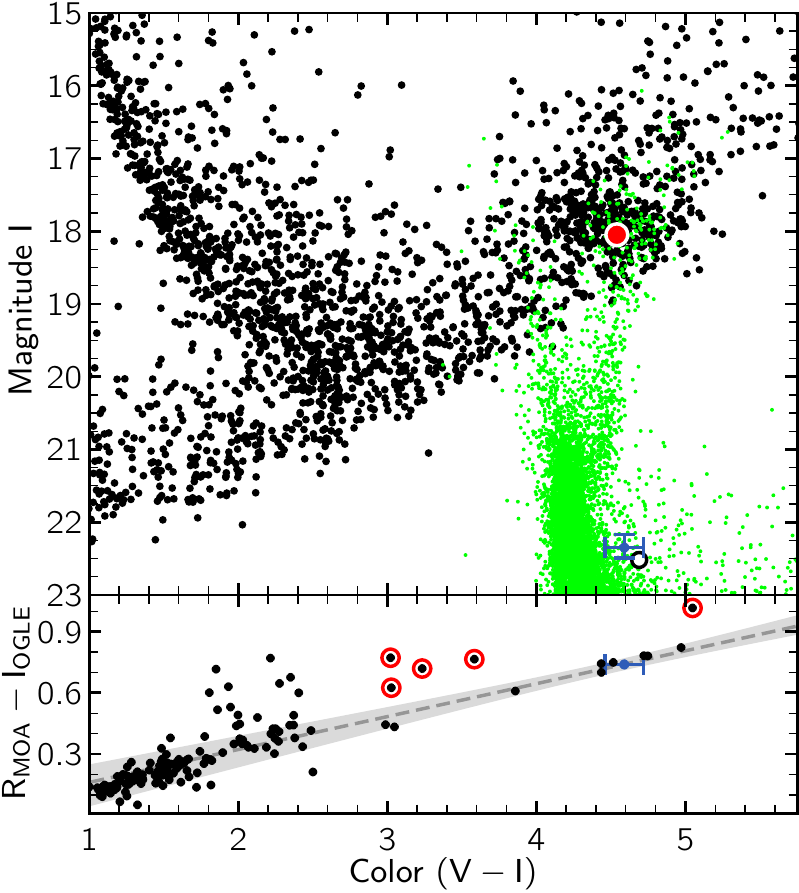}
  \end{center}
  \caption{The upper panel shows the $(V-I, I)$ CMD in the standard Kroni--Cousins $I$ and Johnson $V$
    photometric systems of OGLE-IV stars within $2\arcmin{}$ around
    the source (black dots), not corrected for the interstellar
    extinction. The red circle indicates the RCG centroid,
    the blue dot indicates the source magnitude and color for $s>1$,
    and the black open circle corresponds to the solution $s<1$ (the
    uncertainties are comparable to the case $s>1$). The green dots
    show the \textit{Hubble Space Telescope} CMD from
    \citet{Holtzman1998} shifted to the bulge distance and extinction
    derived in \Sec{sec:rho} for the \ob{15}{1670} line of sight. The
    lower panel shows the empirical color-color transformation between
    the standard photometric system and the instrumental color
    $(\Rmoa-\Iogle)$. The gray shading indicates the 99\% confidence
    interval, and the red circles show the outliers for
    $(V-I) > 3$.}\label{fig:cmd}
\end{figure}

The first step of the source characterization is to calibrate the
instrumental MOA-II magnitudes, $\RK$ and $\VK$, by a
cross-referencing of stars from the MOA-II \textsc{dophot} catalog with
stars in the OGLE-IV catalog. We use these stars to build a catalog
with magnitudes in the standard Kron--Cousins $I$ and Johnson $V$
passbands \citep{Udalski2015}. This calibration is required because
the OGLE-IV field ``BLG500.20'' has not been observed by OGLE-III, and
there was no observation magnified enough in the $V$ band to derive the
source color. A total of 881 stars from the OGLE catalog and within a
2\arcmin{} circle centered on the source are cross-matched with the 167
stars extracted from the same field of view and observed by MOA. From
this, we select stars from the red giant branch to derive the
following relation between the MOA-II instrumental magnitude and the
standard magnitudes and colors \citep{Gould2010..color}:
\begin{equation}\label{eq:color_transfo}
  \RK - I = (0.000 \pm 0.053)
  + (0.161 \pm 0.011) \left(V - I\right)\p
\end{equation}
\Eq{eq:color_transfo} is derived using only the nine
cross-referenced stars found in the red branch in both the MOA and
OGLE catalogs. The instrumental color-color relation, along with the
calibrated OGLE CMD, is shown in \Fig{fig:cmd}.

The CMD plotted in \Fig{fig:cmd} reveals a difference in color of
$\gtrsim 3$ between stars from the red giant branch and the main-sequence stars from the blue plume. It is consistent with a field that
suffers from dust distributed along the line of sight,
with the bluer stars further away from the Galactic bulge than the
redder stars. It is particularly visible when comparing the \textit{Hubble
Space Telescope} ({\itshape HST}) CMD from \citet{Holtzman1998}, shifted to the
extinction of the red clump giant (RCG) in \Fig{fig:cmd}. The event \ob{15}{1670} lies
in a high-extinction region of the Milky Way, at a low $|b|$
($b=-1\fdg12048$), in a field that could be observed by
\textit{WFIRST}.
In the optical $I$ and $V$ passbands, the extinction is more
severe than in the NIR, resulting in a sparse CMD in
\Fig{fig:cmd}, mostly because the brightness in the $V$ passband could not
be measured for many stars.
Extracting the photometry of the
faintest stars is one challenging task, especially in the $V$ band and for
targets with $I \gtrsim 21$.
In particular, the blue stars indicated by the black dots in
\Fig{fig:cmd} and with $I > 21$, are likely suffering from systematic
errors, and we reject them in our analysis.

The next step is to measure the extinction and reddening of stars
close to the source and find its color. We use two independent
methods to find the location of the RCG. On the one
hand, a nonparametric kernel distribution estimation method
identifies a local maximum of the two-dimensional probability
distribution function in the red giant branch due to the RCG stars. This method yields a color
$(V-I)_\mathrm{RCG}=4.51\pm0.15$ and a magnitude
$I_\mathrm{RCG}=17.93\pm0.28$. On the other hand, the centroid of the
RCG stars is $(V-I)_\mathrm{RCG}=4.54\pm0.02$ and
$I_\mathrm{RCG}=18.05\pm0.1$. While the two methods do not provide the
same uncertainties, the results are compatible.
Moreover, we test the reliability of this measurement by searching for
the centroid of the RCG stars located within a $1\arcmin$ circle
(instead of $2\arcmin$) centered on the source. We find
$(V-I)_\mathrm{RCG}=4.54\pm0.03$ and $I_\mathrm{RCG}=18.02\pm0.2$.
These values are well within the error bars of the previous
measurement, thus indicating that the RCG location can be accurately
measured despite the high extinction.

For a source located in the Galactic bulge, the absolute magnitude and
color of the RCG are $M_{I,\mathrm{RCG}}=-0.17\pm0.05$
\citep{Chatzopoulos2015,Nataf2016} and $(V-I)_{\mathrm{RCG},0} = 1.06$
\citep{Bensby2013}. The distance to the RCG can be derived from  the
measurement of the distance to the Galactic center \citep{Nataf2016},
$D_\mathrm{GC}=8.33\,\kpc{}$,
\begin{equation}
  D_\mathrm{RCG} = \frac{D_\mathrm{GC}\sin{(\phi)}}{\cos{(b)}\sin{\left(l+\phi\right)}}\v
\end{equation}
where $\phi=40^\circ$ is the angle between the Galactic bulge major
axis and the line of sight of the Sun. For OGLE-2015-BLG-1670, we find
the RCG to be at a distance of $D_\mathrm{RCG}=8.14\,\kpc{}$,
corresponding to a distance modulus of $\mu=14.55$. If we assume that
the source suffers from the same extinction and reddening as the RCG
(\ie, the source is assumed to be in the Galactic bulge, at $8.14\,\kpc$
from Earth),
the dereddened source magnitude is
$I_\mathrm{s,0}=I_\mathrm{s}+M_{I,\mathrm{RCG}}+\mu-I_\mathrm{RCG}$,
\ie, $I_\mathrm{s,0}=18.68^{+0.20}_{-0.19}$, and for
$(V-I)_\mathrm{s}=4.59^{+0.14}_{-0.13}$, we find
$(V-I)_\mathrm{s,0}=1.11^{+0.14}_{-0.13}$.  These values correspond to
an extinction $A_I=3.67$ (in good agreement with the $A_I=3.5$ derived
from \cite{Gonzalez2012} after the transformation from the NIR to the I band), a
color excess $E(V-I)=3.48$, and a reddening
$R_{V,I}=A_V/E(V-I)=2.05$. In this section, we use the source
brightness and color derived from the solution $s>1$ in
\Tab{tab:model_parameters} in order to explain the method. However, we
include all of the degenerate solutions in the final derivation of the
lens properties (see \Sec{sec:bayes}).

As expected from the visual inspection of \Fig{fig:cmd}, this field
has a high extinction\footnote{For the microlensing event
  KMT-2018-BLG-0073 (Galactic coordinates
  $(l,\,b) = (2.32\arcdeg,\, 0.27\arcdeg)$), \textit{Spitzer} $L$-band
  observations have confirmed a source extinction of $A_I=9.1$ and
  ruled out a scenario with a foreground star superposed on a reddened
  field.}. Despite the difficulty of detecting events at a low
$\left|b\right|$ with optical microlensing surveys, a few events
have already been observed in this region (see
\Sec{sec:intro}). Although the extinction substantially varies at a
subdegree angular scale, we have compared the extinction to the
values derived for \ob{13}{1761}, the closest planetary event
($(l, b) = (0.9368\arcdeg,-1.4842\arcdeg)$). The analysis of this
event yields $E(V-I)=1.87$ and $A_I=1.95$ \citep{Hirao2017}. Although
the extinction is lower, the reddening coefficient $R_{V,I}=2.04$ is
consistent with the value we find. For comparison, in the Baade  
window, \cite{Stanek1996} found a reddening coefficient
$R_{V,I}=A_V/E(V-I)=2.49$, a value broadly consistent with our
measurement despite the higher extinction in the line of sight for
\ob{15}{1670}. Also, from the extinction maps built from the OGLE-III
catalog \citep{Nataf2013}, \ob{15}{1670} lies in a region with
$E(V-I)\geq1.34$, as expected. Finally, for the Galactic coordinates
$(0.5, -1.8)$, extrapolating the empirical law predicting the red
clump magnitude \citep{Nataf2013} beyond its scope, we find a value
$I_\mathrm{RC}=18.18$, consistent with our measurement.

The last step is deriving the angular source size 
from the following empirical relation \citep{Boyajian2014},
\begin{equation}\label{eq:thScolor}
  \log\left(\frac{2\thS}{\mas}\right) = 0.501414 - 0.2 I_\mathrm{s,0}
  + 0.419685 (V-I)_{\mathrm{s},0}\v
\end{equation}
inferred from stars with colors corresponding to
$3900<T_\mathrm{eff}<7000$ \citep{Bennett2017}. We find that the angular
source size $\thS=0.85^{+0.14}_{-0.12}\,\uas$, with error bars
mostly due to the uncertainty on the source color and brightness
rather than the 2\% uncertainty on \eq{eq:thScolor}. The source color
is consistent with a K2--K4 main-sequence star, with an effective temperature
$T_\mathrm{eff}\approx 4600\,\mathrm{K}$.

\begin{table}[tbp]
{\centering
\caption{Lens and Source Properties Derived from the Solutions $s<1$ and $s>1$\\ and
    the Bayesian Analysis Described in \Sec{sec:bayes}}\label{tab:physdeg}
\setlength{\tabcolsep}{8.10pt}
\begin{tabular}{@{} l c c c c @{}}\toprule
Parameter & $s<1$ & $s>1$ & Bayes & Units\\ \midrule
Einstein radius $\thE$                      & $0.392^{+0.077}_{-0.062}$  & $0.395^{+0.084}_{-0.061}$ & $\thEval$                   & $\mas$\\\addlinespace[0.8ex]

Lens-source proper motion $\murelg$         & $5.4^{+1.1}_{-0.9}$        & $6.21^{+1.2}_{-0.95}$     & $\murelgval$                & $\mas\,\yr^{-1}$\\\addlinespace[0.8ex]

Source magnitude\textsuperscript{a} $\Iso$    & $18.85\pm 0.22$            & $18.68^{+0.20}_{-0.19}$   & $18.66\pm0.20$              & \nodata\\\addlinespace[0.8ex]

Source color\textsuperscript{b} $(V-I)_{S,0}$ & $1.21\pm 0.14$             & $1.11^{+0.14}_{-0.13}$    & $1.11\pm0.14$               & \nodata\\\addlinespace[0.8ex]

Source angular radius $\thS$                & $0.87^{+0.16}_{-0.13}$     & $0.85^{+0.14}_{-0.12}$    & $0.85^{+0.14}_{-0.12}$      & $\uas$\\\bottomrule\addlinespace[2ex]
\end{tabular}\par}

{\footnotesize {\bfseries Notes.}

\smallskip\textsuperscript{a}\,$I$-band dereddened source magnitude.

\smallskip\textsuperscript{b}\,Corrected for reddening.}
\end{table}

The combination of the measurement of $\thS$ and \Eq{eq:rho}
yields the Einstein angular radius for the best-fit model,
$\thE=0.395^{+0.084}_{-0.061}\,\mas$. Hence, the lens-source relative
proper motion in the geocentric reference frame is
$\murelg=6.21^{+1.2}_{-0.95} \, \mas\,\yr^{-1}$.
The main results from this section are summarized in
\Tab{tab:physdeg}. This table includes the two degenerate solutions,
$s>1$ and $s<1$, and shows that they yield measurements that are
consistent with each other.

\subsection{Lens Properties}\label{sec:bayes}

\Eq{eq:mass_from_rho} is one relation between the lens mass and
distance. As we could not measure the microlens parallax, the lens
mass cannot be directly derived from the light-curve modeling. However,
all lens configurations are not equally probable. We combine the
microlensing light-curve analysis with a Galactic model in a Bayesian
framework to quantify the relative probability between the different
solutions and find the physical properties of the lens system.


\begin{figure*}[htbp]
{\centering\epsscale{1}
\noindent\begin{minipage}[t]{0.49\textwidth}
\includegraphics[width=\textwidth]{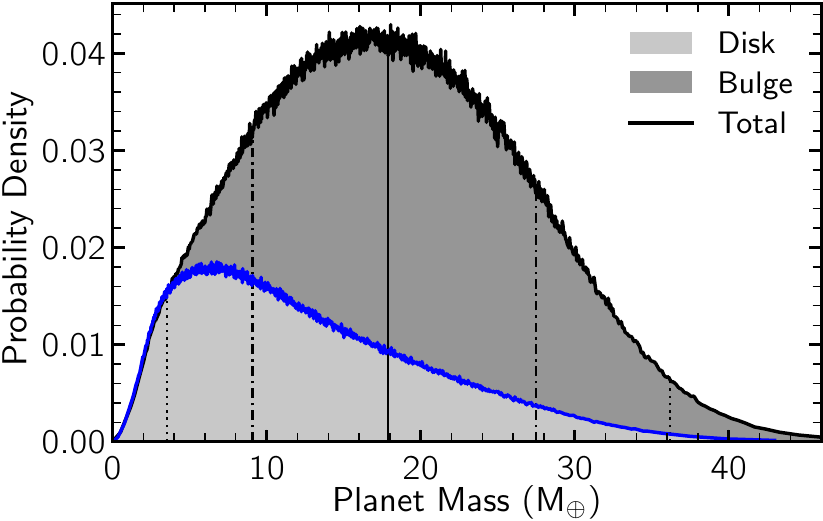}
\end{minipage}\hfill
\noindent\begin{minipage}[t]{0.49\textwidth}
\includegraphics[width=\textwidth]{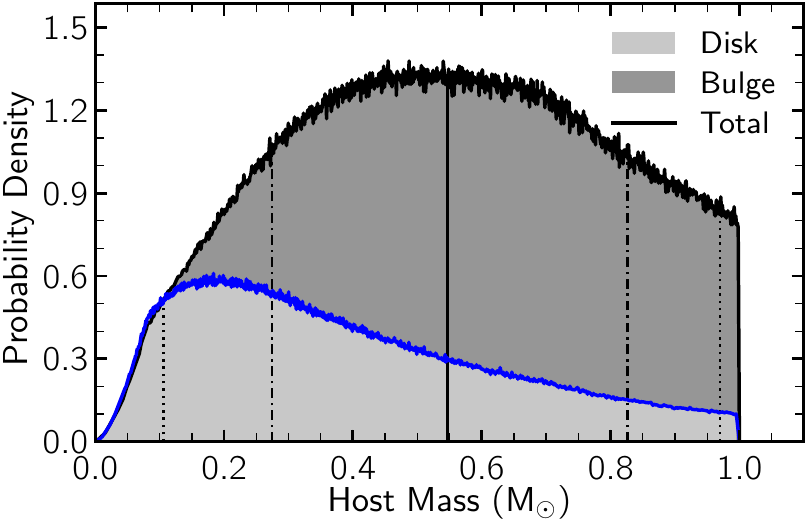}
\end{minipage}

\smallskip\noindent\begin{minipage}[t]{0.49\textwidth}
\includegraphics[width=\textwidth]{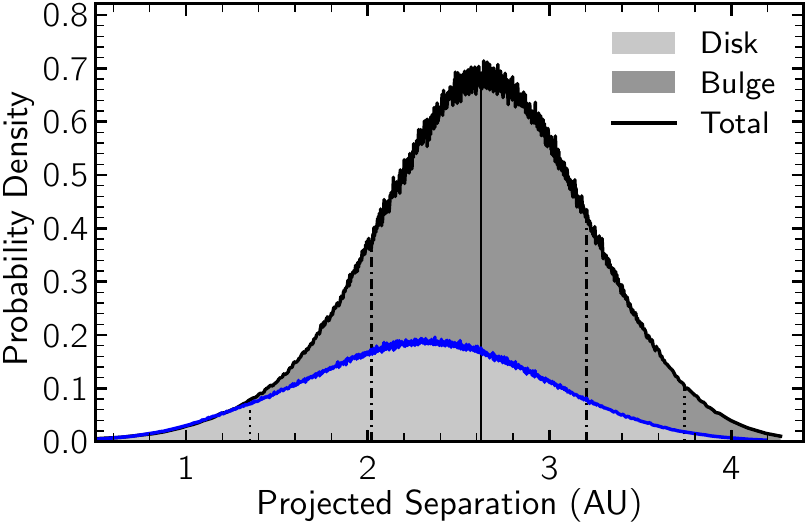}
\end{minipage}\hfill
\noindent\begin{minipage}[t]{0.49\textwidth}
\includegraphics[width=\textwidth]{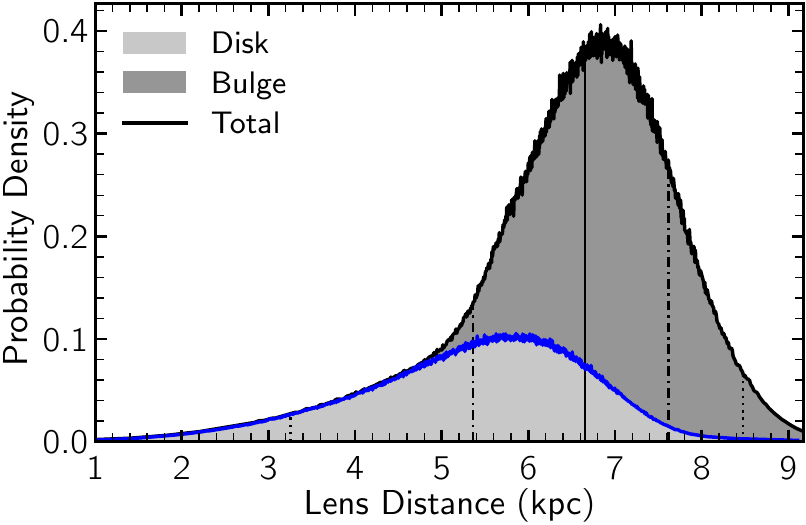}
\end{minipage}}
\caption{Posterior probability distribution of the lens properties
    from a Bayesian analysis that includes the two degenerate
    solutions with $s>1$ and $s<1$ from \Sec{sec:modeling:params},
    weighted by the Galactic model priors described in
    \Sec{sec:bayes}.  Two shaded areas are separated by a blue
    line. They show the contribution of the thin and thick disk (light
    gray), and the spheroid and bulge (dark
    gray) to the posterior distribution (black line). The black
    vertical solid line indicates the median of the distribution, while the
    dotted-dashed and dotted lines respectively show the $68.3\%$ and
    $95\%$ confidence intervals.}\label{fig:bayes}
\end{figure*}

We use the same Galactic model as described in \citet{Bennett2014}
based on stellar densities from \citet{Robin2003} with truncated
escape velocities. This model includes a barred bulge, a spheroid, a
thin disk and a thick disk. This model assumes that, for any given Einstein
radius and mass ratio, the probability for a star to host a planet
does not depend on the host mass. At this stage, we include all
degenerate models found in \Sec{sec:modeling:params} (solutions with
$s>1$ and $s<1$). As shown in \Fig{fig:correlations}, the
posterior probability distributions of each local minimum have similar
statistical properties. Consequently, we weight each Markov chain by
the $\chi^2$ difference between their corresponding best-fit models.

\Fig{fig:bayes} shows the probability distribution of the lens
properties resulting from this Bayesian analysis. As the two
degenerate solutions yield relatively close posterior distributions,
these two solutions do not imply multimodal distributions. As
expected, the lens mass and distance are not well constrained, and the
Galactic priors largely drive the posterior distributions. The source
flux measurement does not exclude main-sequence stars with a mass
larger than $1\,\Msun$, mostly because of the high
extinction. However, such stars are rare in the Galactic bulge, and we
use an upper limit for the lens mass equal to $1\,\Msun$, as shown in
\Fig{fig:bayes}.  The secondary lens component is found to be
$\MLbval\,\Mearth$, which is consistent with a Uranus- or
Neptune-mass planet orbiting a primary lens component with a projected
separation $\aperpval\,\au$. If we assume a circular planetary orbit
with random orientation in space, the three-dimensional orbit radius
is expected to be $\aval\,\au$. This planet is, therefore, orbiting
its host well beyond the snow line. Besides, the host mass derived
from this analysis cannot provide an unambiguous stellar type with an
estimated mass $\MLaval\,\Msun$, consistent with an M dwarf or a
solar-type star. With a lens-source proper motion of
$\murelg = \murelgval \, \mas\,\yr^{-1}$ in the geocentric reference
frame and a lens distance $\DL = \DLval\,\kpc$, the lens may be either
in the disk or in the bulge. In \Fig{fig:bayes}, the light gray shading
indicates the thin and thick disk contribution to the posterior
distribution (black solid curve), while the dark gray shading indicates
the spheroid and bulge contribution. Although these density profiles
raise the possibility of a lens lying in the disk, they also suggest
that a bulge lens is slightly more likely. The results of the Bayesian analysis
are summarized in Tables~\ref{tab:physdeg} and~\ref{tab:bayes}.

\begin{table}[htbp]
{\centering
\caption{Physical Properties of the Lens \ob{15}{1670L} Derived from the Bayesian \nolinebreak Analysis Described in \Sec{sec:bayes}}\label{tab:bayes}
\setlength{\tabcolsep}{12pt}
\begin{tabular}{@{} l c c @{}}\toprule
Parameter & Bayes & Units\\ \midrule
Host mass $M_1$                       & $\MLaval$   & $\Msun$   \\\addlinespace[0.8ex]
Planet mass $M_2$                     & $\MLbval$   & $\Mearth$ \\\addlinespace[0.8ex]
Projected separation $a_\perp$        & $\aperpval$ & $\au$     \\\addlinespace[0.8ex]
Deprojected separation $a$            & $\aval$     & $\au$     \\\addlinespace[0.8ex]
Lens distance $\DL$                   & $\DLval$    & $\kpc$    \\\addlinespace[0.8ex]
Predicted lens magnitude $J_l$        & $\JLvals$   & \nodata   \\\addlinespace[0.8ex]
Predicted lens magnitude $H_l$        & $\HLvals$   & \nodata   \\\addlinespace[0.8ex]
Predicted lens magnitude $K_{s,l}$    & $\KLvals$   & \nodata   \\\bottomrule
\end{tabular}\par}
\end{table}

\section{Summary and Discussion} \label{sec:discussions}

\begin{figure}[tbp]
  \begin{center}
    \includegraphics[scale=1]{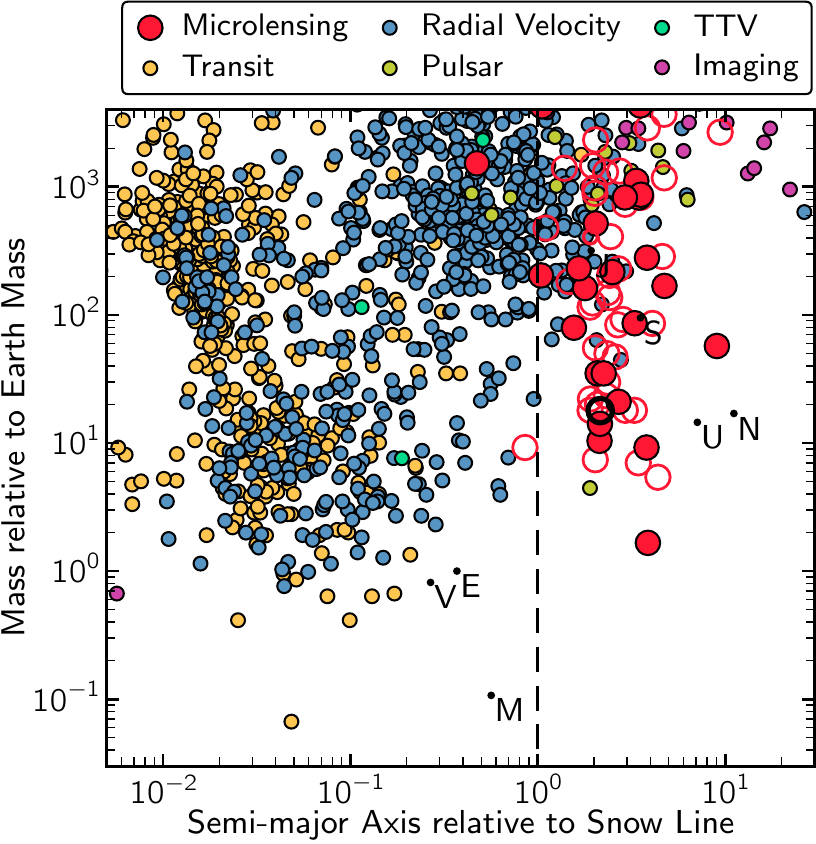}
    \caption{Distribution of known exoplanet masses relative to the
      semi-major axis divided by the snow-line position at $a_\mathrm{snow}
      = 2.7\,\au\,M_1/M_\sun$. Microlensing discoveries with direct host
      star and planet mass measurements are indicated with filled red circles.
      \ob{15}{1670Lb} corresponds to the thick black circle. Each planet
      from our solar system is indicated by its initial (except Mercury).
      Exoplanets not detected using microlensing are from the catalog
      \url{http://exoplanet.eu/} \citep{Schneider2011}.}
    \label{fig:demography}
  \end{center}
\end{figure}

We have presented the analysis of the high-magnification
($A_\mrm{max}\approx230$) microlensing event \ob{15}{1670}.
The anomaly is consistent with a binary lens with a planet-to-host mass
ratio of $q\approx 10^{-4}$. There are two solutions to the event. The best has a
planet-to-host mass ratio $q=1.00^{+0.18}_{-0.16}\times 10^{-4}$ and a projected separation
$s=1.0556^{+0.015}_{-0.0087}$. The second solution has $q=1.50^{+0.39}_{-0.35}\times 10^{-4}$ and
$s = 0.9650\pm 0.0050$ but is disfavored by $\Delta\chi^2=10$.
While we did not detect any reliable parallax signal in the light
curve, the source caustic crossing constrains the angular
source size, $\rho$, in Einstein units. Building the CMD from stars
close to the target, we measured the RCG position and derived the
dereddened source magnitude for the $s>1$ solution, $I_{s,0}=18.68^{+0.20}_{-0.19}$, and
color, $(V-I)_{s,0}=1.11^{+0.14}_{-0.13}$, as well as an estimation of the
source angular size, $\thS=\thSval\,\uas$. The source
size serves as a ``length calibration ruler'' and yields the Einstein
angular radius, $\thE = 0.395^{+0.084}_{-0.061}\,\mas$. The values for the
$s<1$ solution are similar (see \Tab{tab:physdeg}).

This lens mass ratio is very close to the break and the possible peak
in the mass-ratio function identified recently for the first time
\citep{Suzuki.2016} after combining MOA survey observations with
previous statistical investigations \citep{Gould2010, Sumi2010, 
 Cassan.2012} to build the largest sample of microlensing planets in a
study of the planets' demography. For a mass ratio $q<q_\mrm{br}$, the
planet frequency is rising as
$\diff{2}{N}/(\diff{}{\log{q}}\times \diff{}{\log{s}}) = 0.95\times (q
/ q_\mrm{br})^{2.6}\, s^{0.46}$, whereas for $q_\mrm{br}<q$, the
planet frequency is dropping as
$\diff{2}{N}/(\diff{}{\log{q}}\times \diff{}{\log{s}}) = 0.95\times (q
/ q_\mrm{br})^{-0.85}\, s^{0.46}$, where
$q_\mrm{br}=0.67^{+0.90}_{-0.18}\times 10^{-4}$ is the mass-ratio
function break that translates into 1~Neptune mass
($M\approx20\Mearth$) by assuming that M dwarfs dominate the
microlensing planet host sample.  A similar peak in the mass function
around $M=6\,\Mearth$ has been identified in a sample of
\textit{Kepler} planets orbiting M dwarfs (host stars that dominate
the microlensing planet sample) detected by \textit{Kepler}, but for
shorter-period orbits \citep{Dressing2015}. A recent exploration of
the low-mass end of the mass-ratio function has also confirmed the
turnover in the microlensing planet mass function
\citep{Udalski2018}. However, the exact value of the mass-ratio break
$q_\mrm{br}$ is not well constrained due to a lack of planet
detections in the regime $q<q_\mrm{br}$. In this respect,
\ob{15}{1670L} is a noteworthy detection that will tighten constraints
on the lower end of the mass-ratio function.

The measurement of $\thE$ only partially solves the lens mass--distance
degeneracy. However, it is possible to infer the lens physical
properties by conducting a Bayesian analysis that combines the
light-curve modeling with priors on the lens-source relative proper
motion from a Galactic model. The resulting lens consists of a
$\MLbval\,\Mearth$ Neptune-mass planet orbiting a $0.55\pm0.28$ main-sequence star with a projected orbital separation
$\aperpval\,\au$. \ob{15}{1670Lb} is shown in \Fig{fig:demography} as
a thick black circle, together with the distribution of known
exoplanets in mass versus semi-major axis divided by the location of
the snow line, $\asnow$. The location of the snow line in a
protoplanetary disk depends on many parameters, including the host star
properties (age, effective temperature, mass) and its environment
\citep[dust, gas, disk; \eg, see][]{Ida2005, Kennedy2008,
  Min2011}. Its dependency with the host star mass is often assumed to
be a power law and scaled to its current position in the solar System:
$\asnow = 2.7 \, \au \, (M/M_\odot)^\alpha$, with $\alpha=2$ for main-sequence stars whose mass is $0.2 \, \Msun < M < 1.5 \, \Msun$ and
optically thin disks \citep{Ida2005}, or in the range
$[6/9 \mathbin{;} 8/9]$ \citep{Kennedy2008} for hosts with
$M<3 \, \Msun$, depending on the accretion rates and model
assumptions.  For consistency with previous articles reporting new
microlensing detections, we adopt a linear law, \ie, $\alpha=1$. In
\Fig{fig:demography}, exoplanets with a direct mass measurement are
indicated by filled red circles, whereas open red circles show the planets
whose masses have been derived from Galactic models. \ob{15}{1670Lb}
lies well beyond the snow line.

\begin{figure}[tbp]{\centering
    \includegraphics[width=\textwidth]{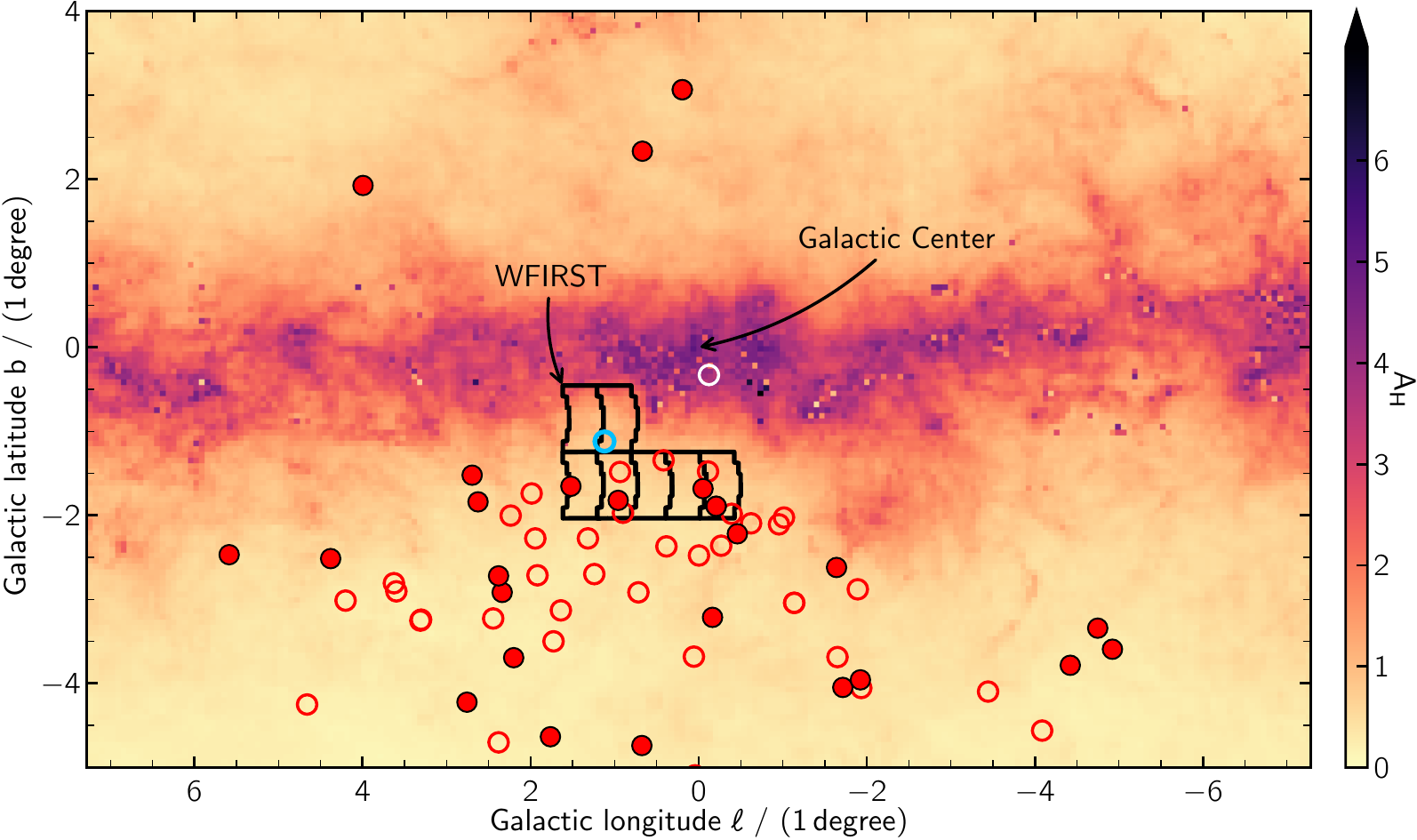}\par}
    \caption{Known exoplanets and brown dwarfs close to the Galactic
      center line of sight. Microlensing detections with direct host
      star and planet mass measurements are indicated with filled red
      circles, while the open red circles correspond to objects with a mass
      estimate. The white circle shows the location of
      UKIRT-2017-BLG-001Lb \citep{Shvartzvald.2018.ukirt}. The background is
      the extinction map in the $H$ passband from
      \cite{Gonzalez2012}. The solid black lines indicate the
      footprints of the seven provisional baseline \textit{WFIRST}-WFI
      fields \citep{Penny.2019.wfirst}, with a total active area of
      $1.96\,\deg^2$. \ob{15}{1670} is shown as a thick blue circle.}
    \label{fig:reddening}
\end{figure}

High-resolution follow-up would help in
measuring the actual mass of the planet in the future, either by
resolving the source and the lens or by a measurement of the excess
flux on top of the source.  Following the same reasoning as in
\Sec{sec:bayes}, we use our Galactic model to predict the lens
brightness in the three passbands. For an extinction $A_J=1.60$,
$A_H=0.99$, and $A_{K_s}=0.65$ \citep{Gonzalez2012}, we estimate the
lens magnitude to be $J_{l}=\JLvalss{}$, $H_l = \HLvalss$, and
$K_{s,l} = \KLvalss$ ($2\sigma$ limits; see \Tab{tab:bayes}
for $1\sigma$ limits). As this event is faint, and we cannot
detect a microlens parallax, the lens brightness remains
uncertain. However, the lens should be bright enough to be observed
from ground-based facilities equipped with adaptive optics (AO), like
Keck, and it will be separated from the source by $42\,\mas$ in about
$7\,\yr$ with a source brightness
$K_{s,\mathrm{source}}\approx18.4\pm0.8$ ($2\sigma$ limits).
Such high-resolution
observations would provide the last missing independent mass--distance
relation.
For example, this method has recently been used successfully to measure
the lens mass of \ob{12}{0950L} after measuring an angular separation between the
source and planetary host of $34\,\mas$ \citep{Bhattacharya.2018}, thanks to
simultaneous high-resolution follow-up images from the \textit{HST}
and the Keck AO system.
It is worth noting that these observations are performed in
the NIR, in passbands that suffer less from the interstellar
extinction.  \Fig{fig:reddening} shows the distribution of the
exoplanets projected on the sky plane in the vicinity of the Galactic
center line of sight. The background is an extinction map in the
H passband, and the black lines show the footprints of the seven
baseline \textit{WFIRST} fields in Galactic coordinates, chosen from
the current best estimates of the microlensing event rates
\citep{Penny.2019.wfirst}. To our knowledge, \ob{15}{1670} is the
planetary event with the lowest absolute Galactic latitude
$\left|b\right|$ discovered by optical surveys and falls in one
provisional \textit{WFIRST} field. The giant planet
UKIRT-2017-BLG-001Lb \citep[white circle in
\Fig{fig:reddening};][]{Shvartzvald.2018.ukirt} has been detected by the NIR
UKIRT microlensing survey at an even lower latitude in the Galactic
bulge.  In these fields, the high stellar density makes less unlikely
events with a source lying in the Galactic disk. Thus, these
detections are important to build a more comprehensive picture of the
low $|b|$ microlensing fields, where the source distance is more
uncertain.
Excess extinction and uncertain source distance both may affect the
accuracy of the lens mass measurement. The full characterization of
\ob{15}{1670} enabled by high-resolution observations would be an
additional illustration of one mass measurement method on which the \textit{WFIRST}
microlensing survey will rely.

As we did not measure the microlens parallax, we could not derive the
distance to the lens. However, the value of the lens-source proper motion,
$\murelgval\,\mas\,\yr^{-1}$, does not rule out a scenario with a lens
and source lying in the Galactic bulge \citep{Kozlowski2006}.  If it
is confirmed that the new exoplanetary system \ob{15}{1670L} lies in
the Galactic bulge, then it will be one more object in the growing
list of planets orbiting stars in the bulge, similar to
MOA-2011-BLG-293Lb \citep{Yee2012,Batista2014..293}, \ob{15}{0051Lb}
\citep{Han2016b}, \ob{14}{1760Lb} \citep{Bhattacharya.2016},
\ob{12}{0724Lb} \citep{Hirao2016}, and \ob{13}{1761Lb}
\citep{Hirao2017}. In the future, it will be possible to use this
sample to assess the planet demography close to the Galactic center
and test whether or not there is a lack of planets in the Galactic
bulge \citep{Penny.2016}.

Ultimately, the upcoming top-ranked mission
from the 2010 Decadal Survey, \textit{WFIRST}, will provide enough
detections along the Galactic bulge line of sight to tightly
constrain not only the mass function of exoplanets beyond the snow line but
also the distance distribution of planets toward the Galactic
bulge. \textit{WFIRST}'s space microlensing survey will have
sensitivity down to the mass of Mars, and it will detect Earths over a
much wider range of separations than ground-based surveys can.

Although for a fraction of events, \textit{WFIRST} will make use
of the microlens parallax to measure the lens masses and distances
\citep[\eg,][]{Refsdal1966, Gould2013..geopiE,Gould2014..1dpiE, Yee2015..piEflux,
  Mogavero2016, Bachelet.2018}, alone or together
with observations from the ground or possibly from the ESA
\textit{Euclid} space telescope \citep{Beaulieu2010,Laureijs2011, 
  Penny2013},
{\it WFIRST}'s main mass measurement channel
will be the high angular resolution.
Indeed, observations from several microlensing seasons from space enable
the direct measurement of the host star flux and the magnitude and
direction of the lens-source relative proper motion
\citep{Bennett2002..space}. The combination of the lens flux with the
lens-source relative proper motion ensures the correct identification
of the host star in the crowded fields toward the Galactic center
\citep{Bhattacharya.2017, Koshimoto.2017} and provides
a direct mass measurement of both the host star and
the exoplanet.
This mass measurement method that will be
employed with \textit{WFIRST} has already been successfully used with
the \textit{HST}
\citep{Bennett.2006.hst235,Bennett.2015.hst169,Batista.2015,Bhattacharya.2018}.
However, an uncertain source distance may affect the accuracy of these
methods.
A proper-motion measurement
allows the calculation of $\thE$, but, as seen in \eq{eq:mass_from_rho}, extracting a
mass--distance relation for the lens still requires assuming the distance to
the source.
As the provisional \textit{WFIRST} survey fields are very
close to the Galactic plane, a source lying within the disk is more
likely than for larger absolute values of the Galactic latitude,
$|b|$, because the stellar density is higher for a line of sight along
the Galactic plane.  Besides, regions at low $\left|b\right|$ suffer
from more extinction. Excess extinction and uncertain source distance
both may affect the accuracy of the lens mass measurement. As a
consequence, the study of low-$\left|b\right|$ events similar to
\ob{15}{1670} with high-resolution follow-up is of prime interest to
develop the \textit{WFIRST} primary mass measurement method and 
investigate the potential trade-off between a higher lensing rate at
low $|b|$ and difficulty in determining the masses.  The NIR
microlensing survey with UKIRT \citep{Shvartzvald.2018.ukirt} is an
example of observations that, together with future NIR surveys,
enable the first measurement of the microlensing event rate in a
passband (and field of view) that overlaps with {\it WFIRST}
specifications. This makes it possible to optimize the overall {\it
  WFIRST} microlensing survey's yield, which
can have a major impact on planet formation theories, planet
demography, and the potential effect of the Galactic environment on
planetary formation.

\acknowledgments

C.R. is grateful to M.~T.~Penny for providing him with the footprints of
the provisional baseline {\itshape WFIRST}-WFI fields\footnote{Field coordinates
  available at
  \url{https://github.com/mtpenny/wfirst-ml-figures/blob/master/fields/wfirst-fields.pdf}.}
from \cite{Penny.2019.wfirst}.
This research has made use of the KMTNet system operated by the Korea
Astronomy and Space Science Institute (KASI), and the data were
obtained at three host sites of CTIO in Chile, SAAO in South Africa,
and SSO in Australia.
The MOA project is supported by JSPS KAKENHI grant Nos.
JSPS24253004, JSPS26247023, JSPS23340064, JSPS15H00781, JP16H06287, and
JP17H02871.
The OGLE project has received funding from the National Science Centre,
Poland, grant MAESTRO 2014/14/A/ST9/00121, to A.U.
The work by C.R. was supported by an appointment to the NASA
Postdoctoral Program at the Goddard Space Flight Center, administered
by USRA through a contract with NASA. D.P.B., A.B., and C.R. were supported
by NASA through grant NASA-80NSSC18K0274.  Work by C.H. was supported
by grant 2017R1A4A1015178 of the National Research Foundation of
Korea.  Work by A.G. was supported by AST-1516842 from the US NSF. A.G.
received support from the European Research Council under the European
Union's Seventh Framework Programme (FP 7) ERC Grant Agreement
No. [321035]. I.G.S. and A.G. were supported by JPL grant 1500811.

\software{Astropy \citep{Astropy.2018}, GetDist \citep{Lewis2002}, Matplotlib \citep{matplotlib}, MuLAn (C.~Ranc \& A.~Cassan, in preparation), NumPy \citep{Oliphant2015}, SciPy \citep{scipy}.}

\bibliography{references}



\end{document}